\documentclass[longauth]{aa}

\usepackage{graphicx,amsmath,amssymb,aasmacros, multirow}
\usepackage{rotating, url, hyperref}
\usepackage[switch]{lineno}
\usepackage{color}

\makeatletter
\renewcommand*{\@fnsymbol}[1]{\ifcase#1\or$\dagger$\else\@arabic{#1}\fi}

\begin{document}

\authorrunning{The H.E.S.S. Collaboration}
\titlerunning{H.E.S.S. observations of microquasars}
\makeatother

\author{\tiny H.E.S.S. Collaboration
\and H.~Abdalla \inst{1}
\and A.~Abramowski \inst{2}
\and F.~Aharonian \inst{3,4,5}
\and F.~Ait Benkhali \inst{3}
\and A.G.~Akhperjanian \inst{6,5}
\and E.O.~Ang\"uner \inst{7}
\and M.~Arrieta \inst{15}
\and P.~Aubert \inst{24}
\and M.~Backes \inst{8}
% ...
}

\institute{
Centre for Space Research, North-West University, Potchefstroom 2520, South Africa \and 
Universit\"at Hamburg, Institut f\"ur Experimentalphysik, Luruper Chaussee 149, D 22761 Hamburg, Germany \and 
Max-Planck-Institut f\"ur Kernphysik, P.O. Box 103980, D 69029 Heidelberg, Germany \and 
Dublin Institute for Advanced Studies, 31 Fitzwilliam Place, Dublin 2, Ireland \and 
National Academy of Sciences of the Republic of Armenia,  Marshall Baghramian Avenue, 24, 0019 Yerevan, Republic of Armenia  \and
Yerevan Physics Institute, 2 Alikhanian Brothers St., 375036 Yerevan, Armenia \and
Institut f\"ur Physik, Humboldt-Universit\"at zu Berlin, Newtonstr. 15, D 12489 Berlin, Germany \and
University of Namibia, Department of Physics, Private Bag 13301, Windhoek, Namibia \and
GRAPPA, Anton Pannekoek Institute for Astronomy, University of Amsterdam,  Science Park 904, 1098 XH Amsterdam, The Netherlands \and
Department of Physics and Electrical Engineering, Linnaeus University,  351 95 V\"axj\"o, Sweden \and
Institut f\"ur Theoretische Physik, Lehrstuhl IV: Weltraum und Astrophysik, Ruhr-Universit\"at Bochum, D 44780 Bochum, Germany \and
GRAPPA, Anton Pannekoek Institute for Astronomy and Institute of High-Energy Physics, University of Amsterdam,  Science Park 904, 1098 XH Amsterdam, The Netherlands \and
Institut f\"ur Astro- und Teilchenphysik, Leopold-Franzens-Universit\"at Innsbruck, A-6020 Innsbruck, Austria \and
School of Chemistry \& Physics, University of Adelaide, Adelaide 5005, Australia \and
LUTH, Observatoire de Paris, PSL Research University, CNRS, Universit\'e Paris Diderot, 5 Place Jules Janssen, 92190 Meudon, France \and
Sorbonne Universit\'es, UPMC Universit\'e Paris 06, Universit\'e Paris Diderot, Sorbonne Paris Cit\'e, CNRS, Laboratoire de Physique Nucl\'eaire et de Hautes Energies (LPNHE), 4 place Jussieu, F-75252, Paris Cedex 5, France \and
Laboratoire Univers et Particules de Montpellier, Universit\'e Montpellier, CNRS/IN2P3,  CC 72, Place Eug\`ene Bataillon, F-34095 Montpellier Cedex 5, France \and
DSM/Irfu, CEA Saclay, F-91191 Gif-Sur-Yvette Cedex, France \and
Astronomical Observatory, The University of Warsaw, Al. Ujazdowskie 4, 00-478 Warsaw, Poland \and
Aix Marseille Universit\'e, CNRS/IN2P3, CPPM UMR 7346,  13288 Marseille, France \and
Instytut Fizyki J\c{a}drowej PAN, ul. Radzikowskiego 152, 31-342 Krak{\'o}w, Poland \and
Funded by EU FP7 Marie Curie, grant agreement No. PIEF-GA-2012-332350,  \and
School of Physics, University of the Witwatersrand, 1 Jan Smuts Avenue, Braamfontein, Johannesburg, 2050 South Africa \and
Laboratoire d'Annecy-le-Vieux de Physique des Particules, Universit\'{e} Savoie Mont-Blanc, CNRS/IN2P3, F-74941 Annecy-le-Vieux, France
% ...
}

\newcommand{\hess}{H.E.S.S.}
\newcommand{\gr}{$\gamma$-ray}

\title{\vspace*{-0.30cm} A search for very high-energy flares from the microquasars GRS 1915+105, Circinus X-1, and V4641 Sgr using contemporaneous \hess\ and \textit{RXTE} observations}

\date{Received 17 November 2015 / Accepted 28 February 2016}

\abstract{Microquasars are potential $\gamma$-ray emitters. Indications of transient episodes of $\gamma$-ray emission were recently reported in at least two systems: Cyg X-1 and Cyg X-3. The identification of additional \gr -emitting microquasars is required to better understand how $\gamma$-ray emission can be produced in these systems.}%
{Theoretical models have predicted very high-energy (VHE) \gr\ emission from microquasars during periods of transient outburst. Observations reported herein were undertaken with the objective of observing a broadband flaring event in the \gr\ and X-ray bands.}%
{Contemporaneous observations of three microquasars, GRS 1915+105, Circinus X-1, and V4641 Sgr, were obtained using the High Energy Spectroscopic System (H.E.S.S.) telescope array and the \ Rossi X-ray Timing Explorer (\textit{RXTE}) satellite. X-ray analyses for each microquasar were performed and  VHE \gr\ upper limits from contemporaneous \hess\ observations were derived.}%
{No significant \gr\ signal has been detected in any of the three systems. The integral \gr\ photon flux at the observational epochs is constrained to be  $I(> 560 {\rm\ GeV}) < 7.3\times10^{-13}$ cm$^{-2}$ s$^{-1}$, $I(> 560 {\rm\ GeV}) <  1.2\times10^{-12}$ cm$^{-2}$ s$^{-1}$, and  $I(> 240 {\rm\ GeV}) < 4.5\times10^{-12}$ cm$^{-2}$ s$^{-1}$ for GRS 1915+105, Circinus X-1, and V4641 Sgr, respectively.}%
{The \gr\ upper limits obtained using \hess\ are examined in the context of previous Cherenkov telescope observations of microquasars. The effect of intrinsic absorption is modelled for each target and found to have negligible impact on the flux of escaping $\gamma$-rays. When combined with the X-ray behaviour observed using \textit{RXTE}, the derived results indicate that if detectable VHE \gr\ emission from microquasars is commonplace, then it is likely to be highly transient.}


\offprints{H.E.S.S.~collaboration,
\\\email{\href{mailto:contact.hess@hess-experiment.eu}{contact.hess@hess-experiment.eu}};
\\$^\star$ corresponding authors
}

\maketitle

\clearpage\newpage

\section{Introduction}\label{sec:intro}

Microquasars are X-ray binaries that exhibit spatially resolved, extended radio emission. The nomenclature is motivated by a structural similarity with the quasar family of active galactic nuclei (AGN). Both object classes are believed to comprise a compact central object embedded in a flow of accreting material, and both exhibit relativistic, collimated jets. In the current paradigm, both microquasars and AGN derive their power from the gravitational potential energy that is liberated as ambient matter falls onto the compact object. Notwithstanding their morphological resemblance, microquasars and radio-loud AGN represent complementary examples of astrophysical jet production on dramatically disparate spatial and temporal scales. Indeed, conditions of accretion and mass provision that pertain to the supermassive ($10^{6}M_{\odot}\lesssim M_{\rm BH} \lesssim 10^{9}M_{\odot}$) black holes that power AGN and of the stellar-mass compact primaries of microquasars are markedly different. 
In the latter, a companion star (or donor) provides the reservoir of matter for accretion onto a compact stellar remnant (or primary), which can be either a neutron star or a black hole. Partial dissipation of the resultant power output occurs in a disk of material surrounding the primary, producing the thermal and non-thermal X-ray emission, which is characteristic of all X-ray binary systems. Microquasars are segregated on the basis of associated non-thermal radio emission, indicative of synchrotron radiation in a collimated outflow, which carries away a sizeable fraction of the accretion luminosity \citep{2004MNRAS.355.1105F}.
In AGN, superficially similar jet structures are known to be regions of particle acceleration and non-thermal photon emission. The resulting radiation spectrum can extend from radio wavelengths into the very high-energy (VHE; $E_{\gamma}>100$ GeV) \gr\ regime. 
Very high-energy \gr\ emission has been observed from many AGN in the blazar sub-class\footnote{ \url{http://tevcat.uchicago.edu/}},
 where the jet axis is aligned close to the observer line-of-sight, as well as from a few radio galaxies (e.g. M87, \citealp{2003A&A...403L...1A}; Cen A, \citealp{2009ApJ...695L..40A}; NGC 1275, \citealp{Aleksic2012}) and starburst galaxies (e.g. M82, \citealp{Acciari2009}; NGC 253, \citealp{Abramowski2012}).

If similar jet production and efficient particle acceleration mechanisms operate in microquasars and AGNs, this might imply that the former object class are plausible sources of detectable VHE \gr\ emission as well, assuming that appropriate environmental conditions prevail. 
The primarily relevant environmental conditions include the density of nearby hadronic material, which provides scattering targets for inelastic proton scattering interactions; these interactions produce pions that produce $\gamma$-rays when they subsequently decay. The ambient magnetic field strength is also important and influences the rate at which electrons lose energy via synchrotron radiation. Synchrotron photons contribute to the reservoir of soft photons that are available for inverse Compton (IC) up-scattering into the VHE $\gamma$-ray regime.
The argument for phenomenological parity between AGN and microquasars, possibly related to their structural resemblance, has been strengthened in recent years as the spectral properties of both radio and X-ray emission are remarkably similar for both stellar mass and supermassive black holes. In recent years these similarities led to the postulation of a so-called fundamental plane, which describes a three-dimensional, phenomenological correlation between the radio (5 GHz) and X-ray (2--10 keV) luminosities and the black hole mass \citep{uniplane, Falcke2004}.
However, the fundamental plane does not appear to extend into the TeV band. To date, only one well-established microquasar has been observed to emit in the VHE \gr\ regime. This is the Galactic black hole Cygnus X-1, which was marginally detected (at the $\sim4\sigma$ level) by the MAGIC telescope immediately prior to a $2-50\,$keV X-ray flare observed by the INTEGRAL satellite, the \textit{Swift} Burst Alert Telescope (BAT), and the \textit{RXTE} All-Sky Monitor (ASM) \citep{2007ApJ...665L..51A,2008A&A...492..527M}. \cite{Laurent2011} recently identified linear polarized soft $\gamma$-ray emission from Cygnus X-1 (see also \citealp{Jourdain2012}), thereby locating the emitter within the jets and identifying their capacity to accelerate particles to high energies (see however \citealp{Romero2014}). 
Further motivation for observing microquasars in the VHE band arises from the recent identification of the high-mass microquasar Cygnus X-3 as a transient high-energy (HE; 100MeV $<E_{\gamma}<$ 100 GeV) \gr\ source by the \textit{Fermi} \citep{2009Sci...326.1512F} and AGILE \citep{2009Natur.462..620T} satellites. The identification of a periodic modulation of the HE signal is consistent with the orbital frequency of Cygnus X-3 and provides compelling evidence for effective acceleration of charged particles to GeV energies within the binary system \citep{2009Sci...326.1512F}. Based on evidence from subsequent reobservations, the HE \gr\ flux from Cygnus X-3 appears to be correlated with transitions observed in X-rays in and out of the so-called ultra-soft state, which exhibits bright soft X-ray emission and low fluxes in hard X-rays and is typically associated with contemporaneous radio flaring activity \cite[e.g.][]{2012MNRAS.421.2947C}. 
Unfortunately, repeated observations of Cygnus X-3 using the MAGIC telescope did not yield a significant detection \citep{2010ApJ...721..843A}, despite the inclusion of data that were obtained simultaneously with the periods of enhanced HE emission detected using \textit{Fermi}. However, the intense optical and ultraviolet radiation fields produced by the Wolf-Rayet companion star in Cygnus X-3 imply a large optical depth for VHE \gr s due to absorption via $e^{+}e^{-}$ pair production \citep[e.g.][]{2010MNRAS.406..689B, 2012MNRAS.421.2956Z}. Accordingly, particle acceleration mechanisms akin to those operating in Cygnus X-3 may yield detectable VHE \gr\ fluxes in systems with fainter or cooler donors.

Mechanisms for $\gamma$-ray production in microquasars have been widely investigated, resulting in numerous hadronic \citep[see e.g.][]{2003A&A...410L...1R} and leptonic \citep[see e.g.][]{1999MNRAS.302..253A, 2002A&A...388L..25G, 2006A&A...447..263B, 2006ApJ...643.1081D, 2010MNRAS.404L..55D} models, describing the expected fluxes and spectra of microquasars in the GeV-TeV band. In both scenarios, a highly energetic population of the relevant particles is required and, consequently, emission scenarios generally localize the radiating region within the jet structures of the microquasar. 
Leptonic models rely upon  IC scattering of photons from the primary star in the binary system or photons produced through synchrotron emission along the jet to produce VHE \gr\ emission. In this latter scenario, they closely resemble models of extragalactic jets \citep{1981ApJ...243..700K,1989ApJ...340..181G}, but typically invoke internal magnetic fields that are stronger by factors $\sim1000$. 
Consideration of hadronic models is motivated by the detection of Doppler-shifted emission lines associated with the jets of the microquasar SS 433 \citep[e.g.][]{1984ARA&A..22..507M}, indicating that at least some microquasar jets comprise a significant hadronic component. 
Models of VHE \gr\ production by hadronic particles generally invoke electromagnetic cascades initiated by both neutral and charged pion decays \citep{2003A&A...410L...1R,1996A&A...309..917A,2005ApJ...632.1093R}. 

Electron-positron pair production, $\gamma\gamma\rightarrow e^{+}e^{-}$ can absorb VHE $\gamma$-rays . In the case of 1 TeV $\gamma$-rays, the cross section for this process is maximised for ultraviolet target photons ($E_{\rm ph}\sim10$ eV), where its value may be approximated in terms of the Thomson cross section as $\sigma_{\gamma\gamma}\approx\sigma_{T}/5$ \citep[e.g.][]{1967PhRv..155.1404G}. In high-mass systems, the companion star is expected to produce a dense field of these target photons to interact with the $\gamma$-rays \citep[e.g.][]{1987ApJ...322..838P,1995SSRv...72..593M,2005ApJ...634L..81B,2006A&A...451....9D}. This process can be very significant and probably contributes to the observed orbital modulation in the VHE \gr\ flux from LS 5039 \citep{2006A&A...460..743A}. In contrast, the ultraviolet spectrum of low-mass microquasars is likely dominated by the reprocessing of X-ray emission in the cool outer accretion flow \citep{1994A&A...290..133V,2009MNRAS.392.1106G}, although jet emission might also be significant \citep{2006MNRAS.371.1334R}. Regardless of their origin, the observed optical and ultraviolet luminosities of low-mass X-ray binaries (LMXBs) are generally orders of magnitude lower than those of high-mass systems \citep{2006MNRAS.371.1334R}, and the likelihood of strong \gr\ absorption is correspondingly reduced.

However, microquasars may only become visible in the TeV band during powerful flaring events. These transient outbursts, characterised by the ejection of discrete superluminal plasma clouds, are usually observed at the transition between low- and high-luminosity X-ray states \citep{2004MNRAS.355.1105F}. Monitoring black-hole X-ray binaries with radio telescopes and X-ray satellites operating in the last decade enabled a classification scheme of such events to be established \citep{Homan2005}. Hardness-intensity diagrams (HIDs) plot the source X-ray intensity against X-ray colour (or hardness) and have subsequently been extensively used to study the spectral evolution of black-hole outbursts. At the transition from the so-called low-hard state to the high-soft states through the hard-to-soft intermediate states, the steady jet associated with the low-hard state is disrupted. These transient ejections, produced once the accretion disk collapses inwards, are more relativistic than the steady low-hard jets \citep{2004MNRAS.355.1105F}. Internal shocks can develop in the outflow, possibly accelerating particles that subsequently give rise to radio optically thin flares observed from black-hole systems;  this phenomenological description is also extensible to neutron stars, although in that case jet radio power is lower by a factor 5--30  (\citealp{2006MNRAS.366...79M}).

Outburst episodes have also been observed in cases in which the source remained in the hard state without transition to the soft state \citep{Homan2005}. The detection (at the $\sim4\sigma$ level) by the MAGIC telescope of the high-mass, black-hole binary Cygnus X-1 took place during an enhanced $2-50\,$keV flux low-hard state as observed with the INTEGRAL satellite, the \textit{Swift} BAT, and the \textit{RXTE} ASM \citep{2008A&A...492..527M}. However, although the source X-ray spectrum remained unchanged throughout the TeV flare, such a bright hard state was unusually long when compared with previous observations of the source.

Here we report on contemporaneous observations with H.E.S.S. and \textit{RXTE} of the three microquasars V4641~Sgr, GRS~1915+105, and Circinus X-1. Information on the targets, the H.E.S.S. and \textit{RXTE} observations, and the corresponding trigger conditions are detailed in Sect.~\ref{sec:targets}. Analysis results are reported in Sect.~\ref{sec:mqs:results} and discussed in Sect.~\ref{sec:mqs:context}. In the appendix, detailed information on the X-ray analysis is reported, which in particular includes HIDs corresponding to the time of observations for the three studied sources.

\section{Targets and observations}
\label{sec:targets} 

\subsection{Observations}\label{sec:mqs:observations}

The \hess\ Imaging Atmospheric Cherenkov Telescope (IACT) array is situated on the Khomas Highland plateau of Namibia (23$^{\circ}16'18\arcsec$ south, $16^{\circ}30'00\arcsec$ east), at an elevation of 1800 m above sea level, and is capable of detecting a Crab-like source close to the zenith at the 5$\sigma$ level within $<5$ minutes under good observational conditions \citep{2006A&A...457..899A}. The point source sensitivity of \hess\ enables it to detect a $2.0\times10^{-13}~\rm{cm}^{-2}\rm{s}^{-1}$ \gr\ flux above 1 TeV, at the 5$\sigma$ level within 25 hours, which, together with a low-energy threshold ($\sim$100 GeV), makes \hess\ an invaluable instrument for studying the VHE $\gamma$-ray emission from microquasars. A fifth and larger telescope (commissioned in 2013) will allow  the energy threshold to be lowered and will further increase the sensitivity of the instrument. For the analysis presented here, H.E.S.S. observations were carried out using the full, original four-telescope array. Owing to the diverse morphologies of the three binary systems, unique observational trigger criteria were established for each target employing various combinations of the observed X-ray state and radio flaring activity. Details are provided in subsequent paragraphs.

The \textit{Rossi X-ray Timing Explorer} (\textit{RXTE}) was a space-based X-ray observatory launched on 30 December 30 1995 and decommissioned on 5 January 2012. The primary mission of \textit{RXTE} was to provide astrophysical X-ray data with high timing resolution. This observatory occupied a circular low-earth orbit with an orbital period of $\sim90$ minutes and carried three separate X-ray telescopes. The \textit{Proportional Counter Array} (PCA) on board \textit{RXTE} comprised five copointing xenon and propane \textit{Proportional Counter Units} (PCUs), which were nominally sensitive in the energy range $\sim2-60$ keV with an energy resolution of $<18\%$ at 6 keV \citep{1993SPIE.2006..324Z}. For studies of rapidly varying sources like X-ray binaries, the PCA timing resolution of $\sim1\;\mu$s can prove invaluable. However, rapid timing measurements also require a bright source to provide sufficient counting statistics within such short time bins. The \textit{High Energy X-ray-Timing Experiment} (HEXTE) comprised two independent clusters of four phoswich scintillation detectors, which were sensitive to photons in the $\sim12-250$ keV energy range and had an energy resolution of $\sim9$ keV at 60 keV. The maximum timing resolution of HEXTE was $\sim8\;\mu$s. The \textit{All-Sky Monitor} (ASM) was a wide field-of-view instrument that monitored $\sim80\%$ of the sky over the course of each $\sim90$ minute orbit. This instrument consisted of three identical scanning shadow cameras and was designed to provide near-continuous monitoring of bright X-ray sources. Nominally, the ASM was sensitive in the energy range from $2-10\,$keV and had a rectangular field of view spanning $110^{\circ}\times12^{\circ}$.

Contemporaneous X-ray (\textit{RXTE}) and VHE \gr\ (\hess) observations were performed at the epochs listed in Table \ref{table:targets}. In the following, we briefly review the observational characteristics of the target microquasars, GRS 1915+105, Circinus X-1, and V4641 Sgr.  Established system parameters that characterise the three target microquasars are collated in Table \ref{tab:target_properties}.

\begin{table*}
\caption{Observationally established parameters of the target microquasars. $P_{\rm orb}$ is the binary orbital period, $M_{\star}$ is the mass of the companion star, $M_{\rm CO}$ is the compact object mass, $\theta_{\rm Jet}$ is the inclination of the observed jet with respect to the line of sight, and $d$ is the estimated distance to the microquasar.}\label{tab:target_properties}
\begin{center}
\begin{tabular}{lllll}
\hline\hline
                                                & GRS 1915+105                  & Circinus X-1               & V4641 Sgr \\
\hline
$P_{\rm orb} $ [d]                      & $33.85\pm 0.16$ (1)                   & 16.6 (2)                   & 2.82 (3)  \\
$M_{\star} [M_{\odot}] $                & $0.47\pm 0.27$ (1)                    &  $3-10$ (5)               & $2.9\pm 0.4$ (10)\\
                                        & $0.28\pm 0.02$ (4)                    &                          &  \\
$M_{\rm CO} [M_{\odot}] $               & $12.4^{+2.0}_{-1.8}$ (6)              &  $\lesssim1.4$ (8)         &  $6.4\pm 0.6$ (10)\\
$\theta_{\rm Jet} [^{\circ}]$           & $60\phantom{.0}\pm 5$ (6)             &  $\lesssim3$   (9)        &  $\lesssim 12$ (3)\\
$d$ [kpc]                               & $8.6^{+2.0}_{-1.6}$ (6,7)             &  $9.4^{+0.8}_{-1.0}$ (9) & $6.2\pm 0.7$ (10)\\
\hline
\end{tabular}
\end{center}
\tablebib{(1) \citet{0004-637X-768-2-185}; (2) \citet{2007ATel..985....1N}; (3) \citet{2001ApJ...555..489O}; (4) \citet{ziolkowski2015}; (5) \citet{1999MNRAS.308..415J, 2007MNRAS.374..999J}; (6) \citet{0004-637X-796-1-2};(7) \citet{Zdziarski2014};(8) \citet{tennant1, 2010ApJ...719L..84L}; (9) \citet{0004-637X-806-2-265}; (10)\citet{0004-637X-784-1-2}; }
\end{table*}

\begin{table*}[th!]
\caption{\label{table:targets} Observational epochs for each target microquasar.}
\centering
\begin{tabular}{ l c c c } \hline \hline

~Target~ & \textit{RXTE} ObsId & ~\textit{RXTE} Observations (MJD)~  & ~H.E.S.S. Observations (MJD)~\\
                             \hline
GRS 1915+105
&90108-01-01-00&53123.091 $\rightarrow$ 53123.109 & 53123.067 $\rightarrow$ 53123.150\\
&90108-01-02-00&53124.074 $\rightarrow$ 53124.094 & 53124.079 $\rightarrow$ 53124.162\\
&90108-01-03-00&53125.130 $\rightarrow$ 53125.149 & 53125.083 $\rightarrow$ 53125.148\\
&90108-01-04-00&53126.114 $\rightarrow$ 53126.129 & 53126.109 $\rightarrow$ 53126.132\\
&90108-01-05-00&53127.097 $\rightarrow$ 53127.114 & 53127.106 $\rightarrow$ 53127.165\\
&90108-01-06-00&53128.150 $\rightarrow$ 53128.165 & 53128.149 $\rightarrow$ 53128.165\\
                             \hline
Cir X-1          
&90124-02-01-00&53174.749 $\rightarrow$ 53174.761 & 53174.748 $\rightarrow$ 53174.832\\
&90124-02-02-00&53175.768 $\rightarrow$ 53175.780 & 53175.735 $\rightarrow$ 53175.822\\
&90124-02-03-00&53176.781 $\rightarrow$ 53176.793 & 53176.772 $\rightarrow$ 53176.858\\
                             \hline
V4641 Sgr        
&90108-03-01-00&53193.904 $\rightarrow$ 53193.924 & Not Observed\\
&90108-03-02-00&53194.887 $\rightarrow$ 53194.908 & 53194.883 $\rightarrow$ 53194.926\\
&90108-03-03-00&53195.871 $\rightarrow$ 53195.892 & 53195.890 $\rightarrow$ 53195.931\\
                             \hline
\end{tabular}
\end{table*}

\subsection{GRS 1915+105}\label{sec:upperlims:grs1915}

GRS 1915+105 is a dynamically established black-hole binary first identified by the WATCH all-sky monitor on board the \textit{GRANAT} satellite \citep{castro}. Observations in the optical and near-infrared using the Very Large Telescope succeeded in identifying the stellar companion as a low-mass KM III giant \citep[][]{greiner2}. GRS 1915+105 gained a measure of celebrity as the prototype Galactic superluminal source \citep{1994Natur.371...46M}.

In a detailed study of the X-ray light curves of GRS 1915+105, \citet[][]{2000A&A...355..271B} succeeded in identifying 12 distinct variability classes, internally characterised by the duration and juxtaposition of three separate spectral states. 
Episodes of class $\chi$ behaviour, belonging to state C and lasting several days, are known as plateaux and are invariably terminated by flaring activity in the radio, infrared, and X-ray bands \citep[][]{2004ARA&A..42..317F}. In contrast with the evidence for self-absorbed synchrotron radiation seen in the spectrally hard, low-luminosity state C, and often associated with continuous relativistic jets \citep[][]{2002MNRAS.331..745K}, radio spectra obtained during the end-plateau flaring episodes indicate optically thin synchrotron emission \citep[][]{1997MNRAS.290L..65F,1998ApJ...494L..61E}. 
Occasionally, these flaring episodes are linked to powerful discrete plasma ejections with instantaneous power output reaching $\gtrsim10^{39}$ erg s$^{-1}$ \citep[][]{1994Natur.371...46M, Zdziarski2014}.
Modelling the emission from these discrete relativistic ejecta, \citet[][]{1999MNRAS.302..253A} showed that inverse Comptonisation of emitted synchrotron photons into the GeV-TeV regime could produce significant and persistent \gr\ fluxes that remain detectable for several days.

\citet{2009A&A...508.1135H} and \citet{2009arXiv0907.1017S} reported VHE \gr\ observations of GRS 1915+105; these authors derived integral flux upper limits of $6.1\times10^{-13}$ cm$^{-2}$s$^{-1}$ above 410 GeV and $1.17 \times10^{-12}$ cm$^{-2}$s$^{-1}$ above 250 GeV, respectively.

For the analysis presented here, GRS 1915+105 was observed by \hess\ between 28 April  and 3 May 2004 in response to an apparent decrease in the 15 GHz radio flux, which was monitored by the Ryle Telescope during a $\sim50$ day plateau state \citep{Pooley2006}, as shown in Figure \ref{evol_GRS1915}, in which coloured markers indicate the \hess\ observation epochs. On the basis of previously observed behaviour, it was thought likely that the observed radio evolution signalled the end of the plateau state and, therefore, that flaring activity would begin within the subsequent 24 hours.
The \textit{RXTE} observations of GRS 1915+105 comprised six individual pointings, contributing to accumulated PCA and HEXTE livetimes of $7.6\,$ksec and $5176\,$s, respectively. Fifteen contemporaneous \hess\ observations were obtained, constituting an overall livetime of 6.9 hours.

\subsection{Circinus X-1}\label{sec:upperlims:cirx1}

Circinus X-1 (hereafter Cir X-1) has been extensively studied since its initial identification \citep{1971ApJ...169L..23M}, revealing a somewhat confusing collection of complex observational characteristics. 

Repeated observation of type I X-ray bursts  \citep[][]{tennant2,tennant1,2010ApJ...719L..84L} definitively identifies the compact primary in Cir X-1 as a low magnetic field ($B\lesssim10^{11}$G) neutron star. Further sub-classification as a Z or atoll source \citep[see, for example,][for an explanation of the distinction between these two classes]{2007A&ARv..15....1D} is not possible since Cir X-1 exhibits several confusing spectral and timing properties, subsets of which are characteristic of both source types \citep[see e.g.][]{1999ApJ...517..472S, ooster}. Accordingly, established paradigms for disk-jet coupling in X-ray binaries with neutron star primaries \citep[e.g.][]{2006MNRAS.366...79M} cannot be reliably employed.

At radio wavelengths, the jets of Cir X-1 display notable structure on arcsecond scales, appearing as a bright core with significant extension along the axial direction of the arcminute jets \citep[][]{1998ApJ...506L.121F}. In fact, the observed extension is rather asymmetric with a ratio of at least two between the observed fluxes of the two opposing jets. Interpreted as pure relativistic aberration, this asymmetry implies a jet velocity $\gtrsim0.1c$. Cir X-1 has also been observed to eject condensations of matter with apparently superluminal velocities $\gtrsim15c$ \citep[][]{fenderradio}. These observations imply a physical velocity for the ejecta $v>0.998c$ with a maximum angle between the velocity vector and the line of sight $\theta<5^{\circ}$.These results identify Cir X-1 as a microblazar,  a Galactic, small-scale analog of the blazar class of AGN, several of which are known sources of VHE $\gamma$-rays. 

Definitive classification of the donor star in Cir X-1 is somewhat problematic. The low apparent magnitude of the detected optical counterpart implies a dereddened luminosity consistent with a low-mass or sub-giant companion, implying that Cir X-1 is a LMXB with a high orbital eccentricity $e\sim0.7-0.9$ \citep[e.g.][]{1999MNRAS.308..415J}. Nonetheless, recent near-infrared \citep[][]{2003A&A...400..655C} and I-band optical \citep[][]{2007MNRAS.374..999J} observations reveal emission features that are consistent with a mid-B supergiant, suggesting a more moderate eccentricity $e\sim0.45$. 

Observations of Cir X-1 in the X-ray band reveal a long-term evolution of the average source brightness. Fluxes rose monotonically from near-undetectable in the early 1970s to a peak value of $\sim 1.5-2$ Crab ($1.5-10\,$keV) at the turn of the millennium, before returning over a period of $\sim4$ years to their pre-rise levels \citep[][]{2003ApJ...595..333P}. Various X-ray spectra, obtained during epochs of both high and low flux, display evidence of complex and variable emission and absorption processes. 

A previous analysis of \hess\ observations of Cir X-1 was presented by \cite{2008AIPC.1085..245N}, who derived a preliminary upper limit to the \gr\ flux above 1 TeV of $1.9\times10^{-13}$ cm$^{-2}$s$^{-1}$  corresponding to a detector livetime of 28 hours. 

The \hess\ observations of Cir X-1 reported here began on 18 June 2004 and were scheduled to coincide with the periastron passage of the binary components. The previous observation of regular radio flares during this orbital interval were thought to provide a good chance of observing during a period of outburst with the associated possibility that superluminal ejections might occur.
The \textit{RXTE} observations of Cir X-1 comprised three individual pointings, corresponding to orbital phase intervals $0.0486\le\phi\le0.0498$, $0.1104\le\phi\le0.1112,$ and $0.1718\le\phi\le0.1725$ (using the radio flare ephemeris of \citet{2007ATel..985....1N}), and contributing to an accumulated PCA livetime of $2576\,$s. A data set comprising 12 contemporaneous \hess\ observations yielded a combined livetime of 5.4 hours.

\subsection{V4641 Sgr}\label{sec:upperlims:v4641sgr}

V4641 Sgr is the optical designation of the habitually weak X-ray source SAX J1819.3-2525 (XTE J1819-254), which was independently identified using the \textit{BeppoSAX} \citep[][]{IAUC7119zand} and \textit{RXTE} \citep[][]{IAUCmarkwardt} satellites. Optical spectroscopic measurements \citep[][]{2001ApJ...555..489O, 2005MNRAS.363..882L} strongly suggest a late B- or early A-type companion with an effective temperature $T_{\rm{eff}}\approx10500$ K. 
The mass of the compact primary, $6.4\pm0.6\, M_{\odot}$ \citep[][]{0004-637X-784-1-2}, categorises V4641 Sgr as a firm black hole candidate. 

V4641 Sgr is probably best known for its exhibition of rapid and violent outbursts. Perhaps the most spectacular of these events was the super-Eddington flare detected by the \textit{RXTE ASM} in September 1999. The observed X-ray fluxes ($2-12\,$keV) increased sharply, reaching $\approx12.2$ Crab within eight hours before fading again to below 0.1 Crab in under two hours \citep[][]{2002A&A...391.1013R}. Powerful contemporaneous flares were also observed at hard X-ray \citep[][]{IAUCmccollough}, optical \citep[][]{IAUC7253stubbings}, and radio \citep[][]{hjellming} wavelengths. In fact, Very Large Array (VLA) radio observations obtained within a day of the X-ray flare resolved a bright jet-like radio structure $\approx0.25$ arcsec in length \citep[][]{hjellming}. Assuming the most likely hypothesis, i.e., that the ejection is coincident with some phase of the X-ray flare, proper motions in the range $0.22\lesssim\mu_{\rm{jet}}\lesssim1.1$ arcsec day$^{-1}$ are derived. At the minimum distance $d=5.5$ kpc, the implied lower limit to the apparent velocity of the ejecta  is $7c\lesssim v_{\rm{min}}\lesssim35c$, which is comparable with the extragalactic jets seen in blazars. Indeed, the remarkably high apparent velocities imply that V4641 Sgr may be a microblazar with a relativistic jet moving close to the line of sight ($\theta_{\rm{jet}}\lesssim12^{\circ}$; from \cite{2001ApJ...555..489O}). Subsequent, weaker broadband outbursts 
have also been observed, suggesting recurrent activity on a timescale $\sim1-2$ years \citep[e.g.][]{hjellming61, 2002IAUC.7928....2R, rupen172, swank295}.

Observations of V4641 Sgr with \hess\ were initiated on 7 July 7 2004 (MJD 53193) in response to the source brightening rapidly in the radio \citep{2004ATel..296....1R}, optical \citep{2004ATel..297....1R}, and X-ray \citep{swank295} bands. The resultant  \textit{RXTE} exposure comprised three observations, each contributing to an accumulated PCA livetime of $5\,$ksec.
Two pairs of  $\sim$30 minute \hess\ observations were obtained contemporaneously with the final two \textit{RXTE} pointings. In total, the four separate exposures constitute an overall livetime of 1.76 hours.

\section{Analysis and results}\label{sec:mqs:results}

X-ray data reduction with the \texttt{FTOOLS 5.3.1} software suite employed the data selection criteria regarding elevation, offset, electron contamination, and proximity to the South Atlantic Anomaly recommended by the \textit{RXTE} Guest Observer Facility website\footnote{http://heasarc.nasa.gov/docs/xte/xhp\_proc\_analysis.html}. For each observation, the PCA STANDARD2 data were extracted from all available PCUs.\ For all observations, HEXTE Archive mode data for both clusters were extracted following the recommended procedures for time filtering and background estimation. Spectral analysis was carried out using the \texttt{XSPEC 12.6.0} package \citep{1996ASPC..101...17A}. Spectral fits for GRS 1915+105 use both PCA and HEXTE data, including an energy range of $3-200\,$keV. For bright X-ray sources, such as GRS 1915+105, statistical errors on the number of counts per spectral bin become insignificant relative to dominant uncertainties in the instrument response. Accordingly, a 1\% systematic error was added to all PCA channels. The remaining sources, Cir X-1 and V4641 Sgr, were not significantly detected by HEXTE and therefore only PCA data in the $3-20\,$keV range were considered to ensure good data quality. 
These targets were sufficiently faint that the spectral bin uncertainties were statistically dominated and the addition of a systematic error component was not required.
In the case of GRS 1915+105, power density spectra (PDS) were derived using the ftool \texttt{powspec}. For each \textit{RXTE} pointing of GRS 1915+105, individual PDS were extracted from $128\,$s intervals comprising $2^{14}$ bins. The resulting spectra were then averaged to produce a PDS for the total light curve with errors estimated using the standard deviation of the average of the power in each frequency bin. The overall PDS were logarithmically rebinned and normalised to represent the squared fractional RMS in each frequency bin \citep[see e.g.][]{1988SSRv...46..273L}. Corrections for instrument deadtime \citep[see, for example,][]{2000A&A...363.1013R} were applied (although this was found to have a negligible effect in the frequency range under consideration) and the expected white noise level was subtracted \citep{1983ApJ...266..160L}. Similar temporal analyses for the remaining targets proved unfeasible because of insufficient count statistics at all but the lowest frequencies.

The \gr\ analysis followed the standard point-source procedure described in \cite{crabpaper}. The reflected background model \citep[see, for example,][]{2007A&A...466.1219B} was used to derive overall results in conjunction with both the \textit{hard} and \textit{standard} event selection cuts described by \cite{crabpaper}. Hard cuts (image size $\ge200$ photoelectrons) tend to enhance the signal of sources with power-law spectral slopes that are harder in comparison to the dominant cosmic ray background. Standard cuts (image size $\ge80$ photoelectrons) provide less sensitivity in such cases but allow a lower energy threshold. No significant detection was obtained for any of the three targets. Upper limits to the VHE $\gamma$-ray flux above the instrumental threshold energy were therefore derived at the 99\% confidence level using the profile likelihood method \citep{2005NIMPA.551..493R}.
 
\subsection{GRS 1915+105}\label{subsec:grs_res}

\begin{figure*}[thp!]
\includegraphics[width=\textwidth]{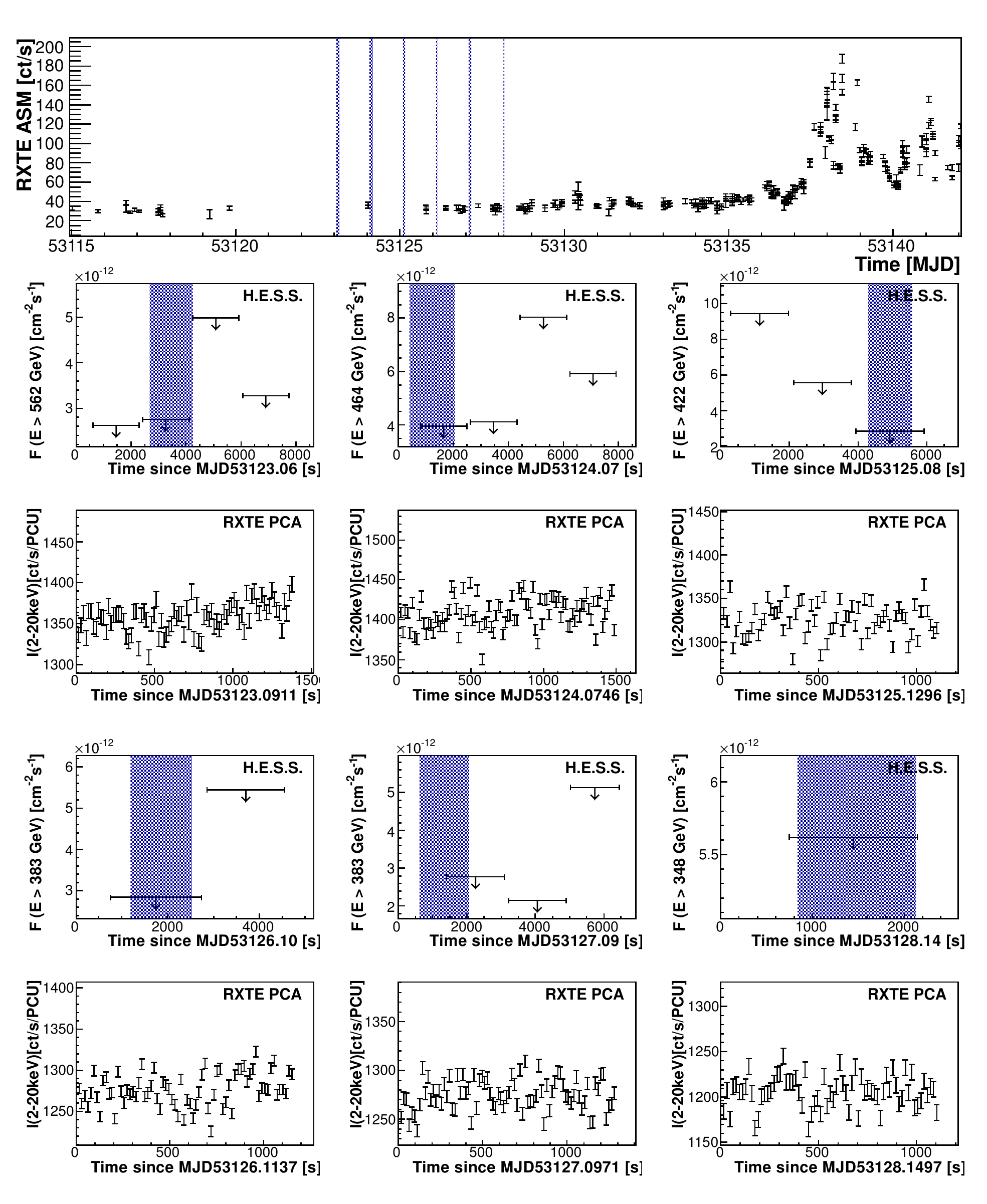}
\caption{\textit{RXTE} ASM, and PCA light curves for GRS 1915+105  together with H.E.S.S. upper limits derived from individual $\sim28$ minute runs using standard event selection cuts. The blue shaded bands on the ASM light curve indicate the extent of the H.E.S.S. observations, while on the H.E.S.S. upper limit plots similar bands illustrate the duration of the contemporaneous PCA observations. The plotted \hess\ upper limits correspond to different threshold energies and the vertical scale of each light curve has been optimised for the plotted data.}\label{fig:grs_lightcurves}
\end{figure*}

As illustrated by the PCA and ASM light curves shown in Figure \ref{fig:grs_lightcurves}, the X-ray count rate was stable to within $\sim10\%$ during each observation and varied by no more than $\sim20\%$ between observations. Indeed, the long-term \textit{RXTE} ASM light curve in Figure \ref{fig:grs_lightcurves} (top panel) clearly indicates that the \hess\ observation epochs occurred during an extended and relatively faint plateau in the $2-10\,$keV flux.

The $3-200\,$keV X-ray spectra shown in Figure \ref{fig:grs1915_spec} also exhibit remarkable stability between observations. The individual spectra are dominated by a hard non-thermal component and strongly suggest class $\chi$ (in state C) behaviour \citep[e.g.][]{2001ApJ...554L..45Z,2001ApJ...558..276T}, which is confirmed by the location of the observations in the HID of Figure \ref{HID_GRS1915}, according to the classification of \citet[][]{2000A&A...355..271B}. 
Figure \ref{evol_GRS1915} shows the contextual X-ray and 15 GHz radio light curves of GRS 1915+105 during a two-month period that brackets the \hess\ observation epochs. It is evident from the figure that \hess\ observed the target during and extended radio-loud plateau ($\sim$80 mJy; for historical flux comparison, a three-year monitoring campaign is presented in \citet{Pooley2006}). The plateau ended approximately ten days later with a combined radio and X-ray flaring episode. The assertion of radio-loud behaviour at the \hess\ observation epochs is supported by the quasi-periodic oscillation (QPO) analysis presented in Figure \ref{fig:grs_powspecs}.
For a detailed discussion see Appendix~\ref{app:GRS1915state}.

In summary, the combined spectral and temporal analyses indicate a robust association of the contemporaneous \hess\ observation with the radio-loud $\chi$ state, and the presence of steady, mildly relativistic jets at the time of observation may be confidently inferred. 

The contemporaneous \hess\ observations did not yield a significant VHE \gr\ detection. The significances corresponding to the total \hess\ exposure are computed using Eq. 17 from \citet{1983ApJ...272..317L} and are listed in Table \ref{tab:mqs:hess_sigmas}. Figure \ref{fig:grs_lightcurves} plots runwise 99\% confidence level upper limits to the integral VHE \gr\ flux above the instrumental threshold energy and illustrates the overlap between the \textit{RXTE} and \hess\ observations. Integral flux upper limits, which correspond to the overall \hess\ exposure, are listed in Table \ref{tab:mqs:hess_uls}. 

\begin{table*}[th!]
\caption{\hess VHE \gr\ significances corresponding to hard and standard event selection regimes.}\label{tab:mqs:hess_sigmas}
\centering
\begin{tabular}{lllllll}
\hline\hline
Target & Image Cuts &$N_{\rm ON}$  [events]& $N_{\rm OFF}$ [events]& $\alpha$ & Excess [events]& Significance [$\sigma$] \\
\hline
\multirow{2}{*}{GRS 1915+105} & Standard & 471 &        7127 &  0.073   & -51.6 & -2.2 \\
        & Hard &  36&   783     &0.060 &        -10.9 & -1.6 \\ 
\hline
\multirow{2}{*}{Cir X-1} & Standard & 385&      5959&   0.068&  -20.1&  -1.0 \\ 
        & Hard & 45     &648    &0.056  &\phantom{-}9.1 &\phantom{-}1.4 \\ 
\hline
\multirow{2}{*}{V4641 Sgr} & Standard & 161     &2373&  0.067&  \phantom{-}1.2  &\phantom{-}0.1\\ 
        & Hard & 11     &275    &0.055  &-4.2   &-1.11\\ 
\hline
\end{tabular}
\end{table*}

\begin{table*}[th!]
\caption{\hess\ VHE \gr\ integral flux upper limits above the telescope energy threshold corresponding to both event selection regimes. The upper limits are derived at the 99\% confidence level, assuming a power-law spectrum ($dN/dE\propto E^{-\Gamma}$) with the photon index $\Gamma_{\rm std} = 2.6$ for standard cuts and $\Gamma_{\rm hard} = 2.0$ for hard cuts. The rather high threshold energies derived for GRS 1915+105 and Cir X-1 are the result of large maximum observational zenith angles.}\label{tab:mqs:hess_uls}
\centering
\begin{tabular}{llllll}
\hline\hline
Target & Cuts & $T_{\rm Live}$ [s] & $\bar{Z}_{\max}$ [$^{\circ}$]  & $E_{\rm thresh}$ [GeV]& $I(>E_{\rm thresh})$ [ph cm$^{-2}$s$^{-1}$] \\
\hline
\multirow{2}{*}{GRS 1915+105} & Standard & 24681 & 40.6 &  562 &$<7.338\times10^{-13}$\\
        & Hard & 24681 & 40.6 & 1101 & $<1.059\times10^{-13}$\\
\hline
\multirow{2}{*}{Cir X-1} & Standard & 19433 & 43.6& 562 &$<1.172\times10^{-12}$ \\
        & Hard & 19433 & 43.6 & 1101 & $<4.155\times10^{-13}$\\
\hline
\multirow{2}{*}{V4641 Sgr} & Standard & 6335 & 8.4 &  237 &$<4.477\times10^{-12}$\\
        & Hard & 6335 & 8.4 & 422 & $<4.795\times10^{-13}$ \\
\hline

\end{tabular}
\end{table*}

\subsection{Cir X-1}\label{subsec:cir_res}

\begin{figure*}[thp!]
\includegraphics[width=\textwidth,angle=0]{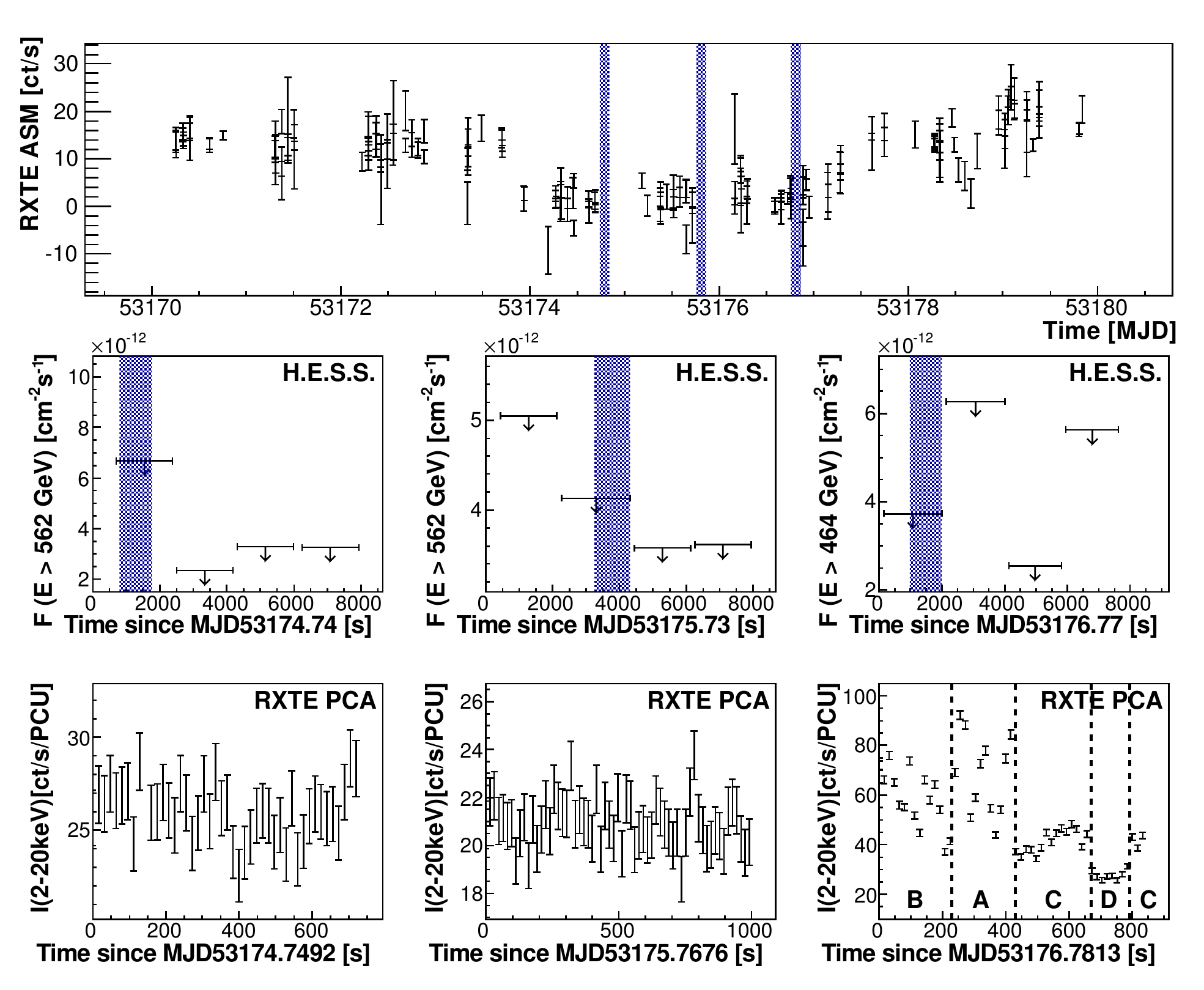}
\caption{\textit{RXTE} ASM and PCA light curves for Cir X-1  together with H.E.S.S. upper limits derived from individual $\sim28$ minute runs using standard event selection cuts. The blue shaded bands on the ASM light curve indicate the extent of the H.E.S.S. observations, while on the H.E.S.S. upper limit plots similar bands illustrate the duration of the contemporaneous PCA observations (OBS $1-3$). The partitioning of OBS 3 into sub-intervals A$-$D based on $2-20\,$keV X-ray flux is illustrated in the bottom right panel. The plotted \hess\ upper limits correspond to different threshold energies, and the vertical scale of each light curve has been optimised for the plotted data.}\label{fig:cir_lightcurves}
\end{figure*}

The ASM light curve shown in Figure \ref{fig:cir_lightcurves} reveals that the \hess\ observation epochs occurred during an extended $\sim4$ day dip in the $2-10\,$keV X-ray flux. Additionally, it should be noted that the observations reported here were obtained during an extremely faint episode in the secular X-ray flux evolution of Cir X-1 \citep{2003ApJ...595..333P}, which is also evident from the HID presented in Figure \ref{HID_CirX1}. As a consequence, the measured X-ray fluxes are significantly lower than most others reported for this source. As illustrated in Figure \ref{fig:cir_lightcurves}, the individual PCA light curves obtained during  the first two pointings are characterised by a relatively low count rate, which remains approximately constant throughout each observation. In marked contrast, the third observation exhibits clear variability with count rates doubling on timescales of $\sim50$s. 

A detailed analysis of the obtained spectra (see Appendix~\ref{app:CirState}) reveals that the observed flux variability is accompanied by marked variations in spectral shape. These can be interpreted as hinting towards a strong mass transfer during the periastron passage and subsequent dramatic evolution of the local radiative environment.

\hess\ observations obtained contemporaneously with the \textit{RXTE} pointings yield a non-detection that is evident from the significances listed in Table \ref{tab:mqs:hess_sigmas}. Figure \ref{fig:cir_lightcurves} plots runwise 99\% confidence level upper limits to the integral VHE \gr\ flux above the instrumental threshold energy and illustrates complete overlap between the \textit{RXTE} and \hess\ observations. Integral flux upper limits, which correspond to the overall \hess\ exposure, are listed in Table \ref{tab:mqs:hess_uls}.

\subsection{V4641 Sgr}\label{subsec:v46_res}
\begin{figure*}[thp!]
\includegraphics[width=\textwidth]{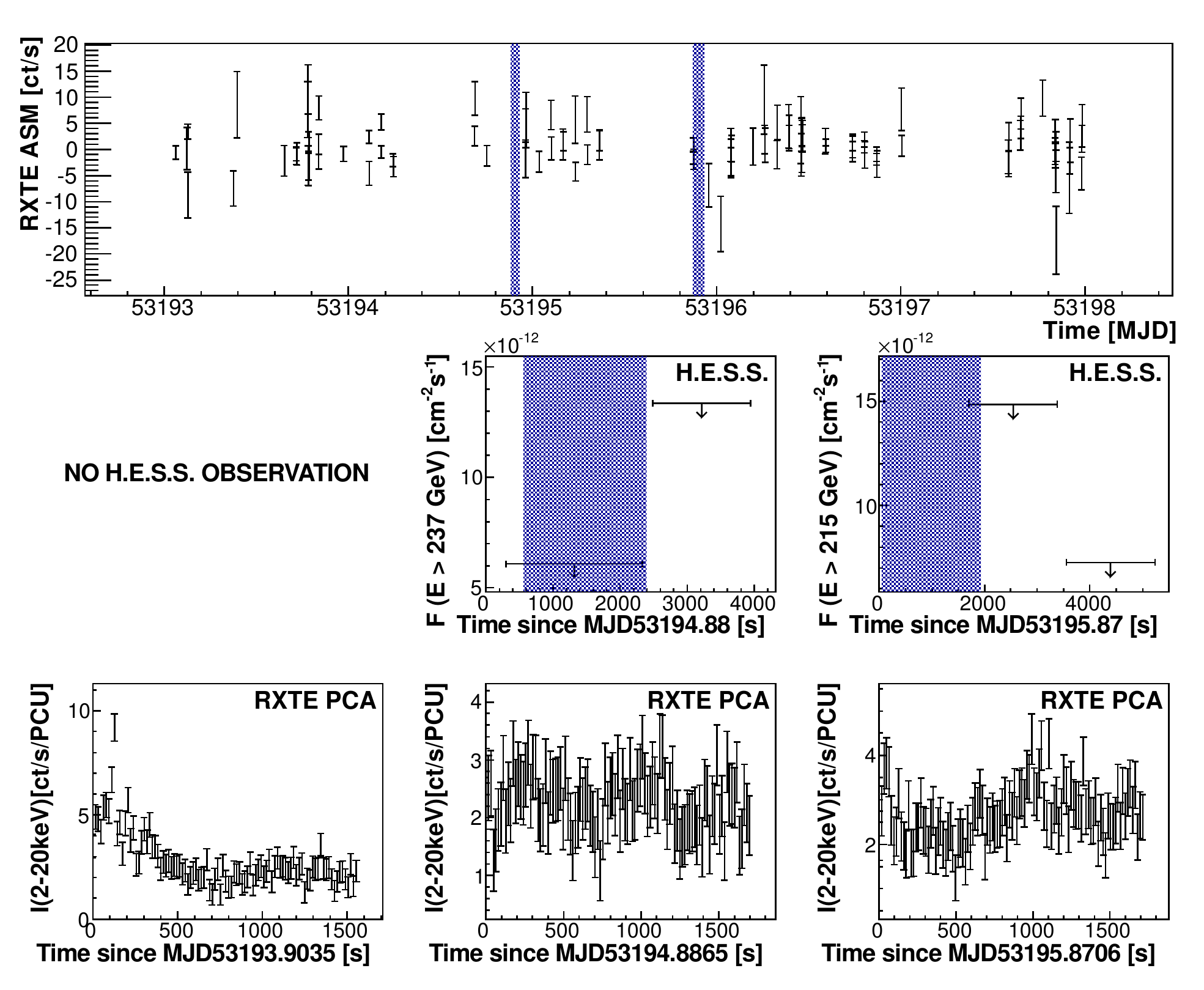}
\caption{\textit{RXTE} ASM and PCA light curves for V4641 Sgr  together with H.E.S.S. upper limits derived from individual $\sim28$ minute runs using standard event selection cuts. The blue shaded bands on the ASM light curve indicate the extent of the H.E.S.S. observations, while on the H.E.S.S. upper limit plots they illustrate the duration of the contemporaneous PCA observations. The plotted \hess\ upper limits correspond to different threshold energies and  the vertical scale of each light curve has been optimised for the plotted data.}\label{fig:v4s_lightcurves}
\end{figure*}
Figure \ref{fig:v4s_lightcurves} shows \textit{RXTE} PCA light curves derived from three pointed observations. The individual light curves indicate various degrees of X-ray variability with the clearest evidence for flaring visible as a sharp $\sim5$-fold count rate fluctuation during the first observation. In marked contrast, the second observation is uniformly faint with the $\chi^{2}$ probability of constant count rate $P_{\rm const}=0.97$ and, hence, this observation is consistent with a period of steady, low-level emission. Subsequently, the third observation reveals a reemergence of mild variability ($P_{\rm const}=0.07$) with $\sim2$-fold count rate fluctuations occurring on timescales of $\sim500$s. 

Radio data shown on Figure \ref{fig:V4641hardness_radio} right were obtained using the VLA and Australia Telescope Compact Array (ATCA) between MJD 53190 and MJD 53208. They indicate rapid variability with peak flux densities of $\sim30$ mJy observed on MJD 53193 \citep{2004ATel..296....1R, 2004ATel..302....1S,2004ATel..303....1R}. An optically thin radio spectrum ($S_{\nu}\propto \nu^{-0.7}$) observed on MJD 53191 was interpreted by \citet{2004ATel..296....1R} as the signature of a decaying radio flare.
Radio observations were triggered by an optical alert from VSNET (MJD 53190) in combination with a \textit{RXTE} PCA measurement during a Galactic bulge scan (MJD 53189) that
 revealed a 2-10 keV X-ray flux equivalent to 8.2 mCrab. For comparison, the August 2003 flare of V4641 Sgr reached 66 mCrab, while quiescent fluxes are typically < 0.5 mCrab \citep{swank295}. 
As shown in Figure \ref{fig:V4641hardness_radio} right, the dedicated \textit{RXTE} PCA observation and \hess\ observations took place between two radio flares, which is consistent with the X-ray variability evolution illustrated in Figure \ref{fig:v4s_lightcurves}.

While V4641 Sgr is evidently the most X-ray-faint binary in the studied sample, it simultaneously exhibits the hardest spectrum, as shown by the hardness values in Figure \ref{fig:V4641hardness_radio} (left-hand panel). Furthermore, the evolution of the hardness is consistent with contemporaneous observations of rapid flux evolution in the radio band \citep{2004ATel..302....1S}\footnote{ \url{http://www.ph.unimelb.edu.au/~rsault/astro/v4641/}} (Figure \ref{fig:V4641hardness_radio}).
To help place the \hess\ and \textit{RXTE} observations in a historical context, the HID for V4641 Sgr in Figure \ref{HID_V4641Sgr} displays the entire archival \textit{RXTE} PCA data set for this target, and compares the X-ray characteristics corresponding to the \hess\ observation periods (different symbols are used to indicate observations obtained on each day in the range MJD 53193-5) with three flaring episodes observed with \textit{RXTE}. 
On  15 September  1999 (orange markers in Figure \ref{HID_V4641Sgr}), a 1500s \textit{RXTE} observation revealed a source flux evolution that is characterised by rapid, large-amplitude variability before reverting to a soft, low intensity state after $\sim1000$s. An optical flare that preceded the \textit{RXTE} observations likely corresponds with the onset of the short 10-hour outburst, which \citet{Wijnands2000} associated with a low $\dot{M}$ accretion event. Historically, flaring episodes exhibited by V4641 Sgr are often short, unpredictable, and relatively faint, which implies that many may go unnoticed. 

Data corresponding to two longer outbursts, spanning the periods 24-26 May  2002 and   5-7 August 2003, are also illustrated in Figure \ref{HID_V4641Sgr}. In the coordinates of the HID, both episodes are topologically similar to the 1999 outburst, but shifted towards fainter harder regions. 

Evidently, the X-ray fluxes that correspond with \hess\ observation epochs indicated in Figure \ref{HID_V4641Sgr} are substantially fainter than any of these historically remarkable outbursts.

In summary, in view of the various multiwavelength data, it seems likely that V4641 Sgr underwent a period of mild activity that spanned the \hess\ observation epochs.

The contemporaneous \hess\ data are consistent with non-detection with the corresponding \gr\ significances listed in Table \ref{tab:mqs:hess_sigmas}. Technical issues prevented \gr\ data corresponding to the first \textit{RXTE} observation from being obtained. 
Simultaneous \gr\ observations were obtained corresponding to the second \textit{RXTE} exposure, which showed no indications of X-ray variability. Although the source began to show increased X-ray activity during the third \textit{RXTE} observation, the degree of overlap with the corresponding \hess\ observations was minimal. At radio, optical, and X-ray energies, V4641 Sgr exhibits rapid variability on timescales $\sim10$ minutes or less \citep[e.g.][]{2005IBVS.5626....1U,2006ApJ...637..992M}. Optimistically, the compelling evidence for mild broadband flaring admits the possibility that the \hess\ observations monitor a transient outburst event.

Integral flux upper limits above the instrumental threshold energy, which correspond to the overall \hess\ exposure at the position of V4641 Sgr, are listed in Table \ref{tab:mqs:hess_uls}.

\section{Discussion}\label{sec:mqs:context}

The principal aim of this investigation was to obtain contemporaneous X-ray and VHE \gr\ observations of three known superluminal microquasars during major flaring events. However, the results presented in $\S$\ref{sec:mqs:results} indicate that the interpretation of the VHE \gr\ non-detections cannot proceed under the assumption of energetic flaring or bulk superluminal ejections at the time of observation. Nonetheless, upper limits to the VHE \gr\ flux were  derived and an analysis of the contemporaneous \textit{RXTE} observations has helped to reveal the X-ray behaviour corresponding to the \hess\ observation epochs. These data facilitate the straightforwards derivation of constraints on the \gr\ luminosity of the target binary systems. In Table \ref {tab:mqs:luminosties} the calculated flux upper limits were used to infer the maximum \gr\ luminosities above the target-specific, instrumental threshold energy for each target binary system by assuming the maximum source distance estimate presented in Table \ref{tab:target_properties}.

\begin{table*}[thp!]
\caption{Estimated maximum VHE \gr\ luminosities of the target microquasars, which would still be consistent with a non-detection given the flux upper limits presented in Table \ref{tab:mqs:hess_uls}. Source distances correspond to the largest estimate that was found in the literature (see $\S$ \ref{sec:targets}).  The energy threshold of Cherenkov telescope arrays increases with observational zenith angle.}\label{tab:mqs:luminosties}
\centering
\begin{tabular}{llll}
\hline\hline
Target & Maximum Distance Estimate& $E_{\rm thresh}$& Luminosity above $E_{\rm thresh}$\\
& [kpc] & [GeV] & [erg s$^{-1}$] \\
\hline
GRS 1915+105 & 10.6 & 562 & $<2.3\times10^{34}$\\
Cir X-1 & 10.2 & 562 & $<3.4\times10^{34}$\\
V4641 Sgr & 6.9 & 237 & $<2.5\times10^{34}$\\
\hline
\end{tabular}
\end{table*}

Analysis of the contemporaneous X-ray and radio observations conclusively places GRS 1915+105 in a radio-loud plateau state at the time of observation. In contrast with the superluminal flaring episodes, this state is characterised by the production of continuous, mildly relativistic radio jets with an estimated power of $\sim3\times10^{38}$ erg s$^{-1}$ (\cite{2002MNRAS.331..745K}, assuming a distance of 11 kpc). Theoretically, it seems unlikely that bright VHE \gr\ emission would be expected from the compact self-absorbed jets, which are typical of the plateau state of GRS 1915+105. For example, a leptonic emission model developed by \cite{2006A&A...447..263B} to simulate the broadband emission of microquasar jets in the low-hard state predicts VHE \gr\ luminosities $\lesssim10^{33}$ erg s$^{-1}$ that are consistent with the \hess\ non-detection. 
Notwithstanding the plausibility of VHE \gr\ emission in the plateau state, a comparison of the estimated jet power with the maximum \gr\ luminosity listed in Table \ref{tab:mqs:luminosties} reveals that the jet power conversion efficiency is constrained to be $\lesssim0.008\%$ for \gr\ production above 562 GeV. For comparison, corresponding efficiency estimates for \gr\ production were derived for the steady, compact jets of other microquasars that were observed in appropriate states. The published MAGIC upper limit on the VHE \gr\ luminosity of Cygnus X-3 during its hard state implies a somewhat larger maximum conversion efficiency of 0.07\% \citep{2010ApJ...721..843A} and a similar value is obtained from MAGIC upper limits on the steady VHE emission from Cygnus X-1 \citep{2007ApJ...665L..51A}.
These efficiencies are inferred from the directly observed jet power, and should be distinguished from the higher jet powers that were indirectly derived from the observation of radio-emitting bubbles inflated by microquasar jets (see e.g. \citealp{Gallo2005} for Cyg X-1, and \citealp{Pakull2010}, \citealp{Soria2010} for S26 in NGC 7793).

We presented an analysis of the entire \hess\ data set for GRS 1915+105  \citep{2009A&A...508.1135H} and we derived an upper limit to the \gr\ flux above 0.41 TeV of $6.1\times10^{-13}$ cm$^{-2}$s$^{-1}$, corresponding to a detector livetime of 24.1 hours. The somewhat higher upper limits presented in $\S$\ref{subsec:grs_res} utilise a more limited data set and are therefore consistent with the previously published value. None of the \hess\ observations of GRS 1915+105 coincide with bright flaring episodes at longer wavelengths.

Observations of Cir X-1 were obtained during an extended dip in the X-ray flux, at phase intervals close to the periastron passage of the binary components. Spectral analysis of the \textit{RXTE} data showed some evidence for a recent increase in mass transfer, producing strong signatures of X-ray absorption. It was hoped that \hess\ observations would coincide with one of the quasi-regular radio flares, which often accompany periastron passage in Cir X-1. 

The ephemeris of \citet{2007ATel..985....1N} predicts the onset of a radio flare $\sim19-20$h before the first \textit{RXTE} observation. Unfortunately, despite the undoubted occurrence of quasi-periodic radio flares from Cir X-1 near periastron, a robust correlation between the observed X-ray and radio behaviour is yet to be identified. Although rapid brightening of the X-ray continuum might indicate accompanying radio flares, evidence for a definitive association is far from clear \citep{2007wmdr.confE..37S, 2008MNRAS.390..447T}. Recent radio observations of Cir X-1 \citep[e.g.][]{fenderradio, 2008MNRAS.390..447T} focus primarily on the ultrarelativistic ejection events that manifest as $\gtrsim3$ day episodes of flaring on timescales of a few hours. In principle, the lack of contemporaneous radio data admits the possibility of such persistent outbursts at the time of observation. By analogy with canonical black hole binaries, it is possible that the inferred variation in the mass accretion rate between the first and second \textit{RXTE} observations also implies an evolution of the jet properties \citep{2006MNRAS.366...79M}, but this is far from clear in such an unusual system. Moreover, \citet{2008MNRAS.390..447T} report compelling evidence that prior to 2006, Cir X-1 underwent a $\sim6$ year episode of unusual radio quiescence, suggesting that jet formation was somewhat suppressed during the epochs of \hess\ observation. 
Accordingly, without strictly simultaneous radio data indicating otherwise, the most likely scenario is that no outflows were present. In this context the absence of a detectable \gr\ signal is not surprising.   

As a confirmed high-mass black hole binary, V4641 Sgr is the studied target that most closely resembles the Cygnus X-1 and Cygnus X-3 systems. Moreover, the \hess\ observations were obtained during a period of sporadic broadband flaring, and comparing these observations with the results of \cite{2007ApJ...665L..51A}, VHE \gr\ emission might have been expected. The detection of Cyg X-1 using the MAGIC telescopes appeared to coincide with the rising part of a strong X-ray flare. In contrast, radio spectra obtained close to the \hess\ observational epochs are indicative of the decay following a flaring episode \citep{2004ATel..302....1S}. 
Assuming that the \gr\ emission mechanisms operating in Cyg X-1 also occur in V4641 Sgr, the absence of a significant \hess\ detection might be viewed as evidence that production of GeV and TeV photons is a highly transient process. This would further suggest that \gr\ emission originates in a spatially compact region that is at most a few light hours in size.

\begin{figure}[h!]
\begin{center}
\includegraphics[width=0.53\textwidth]{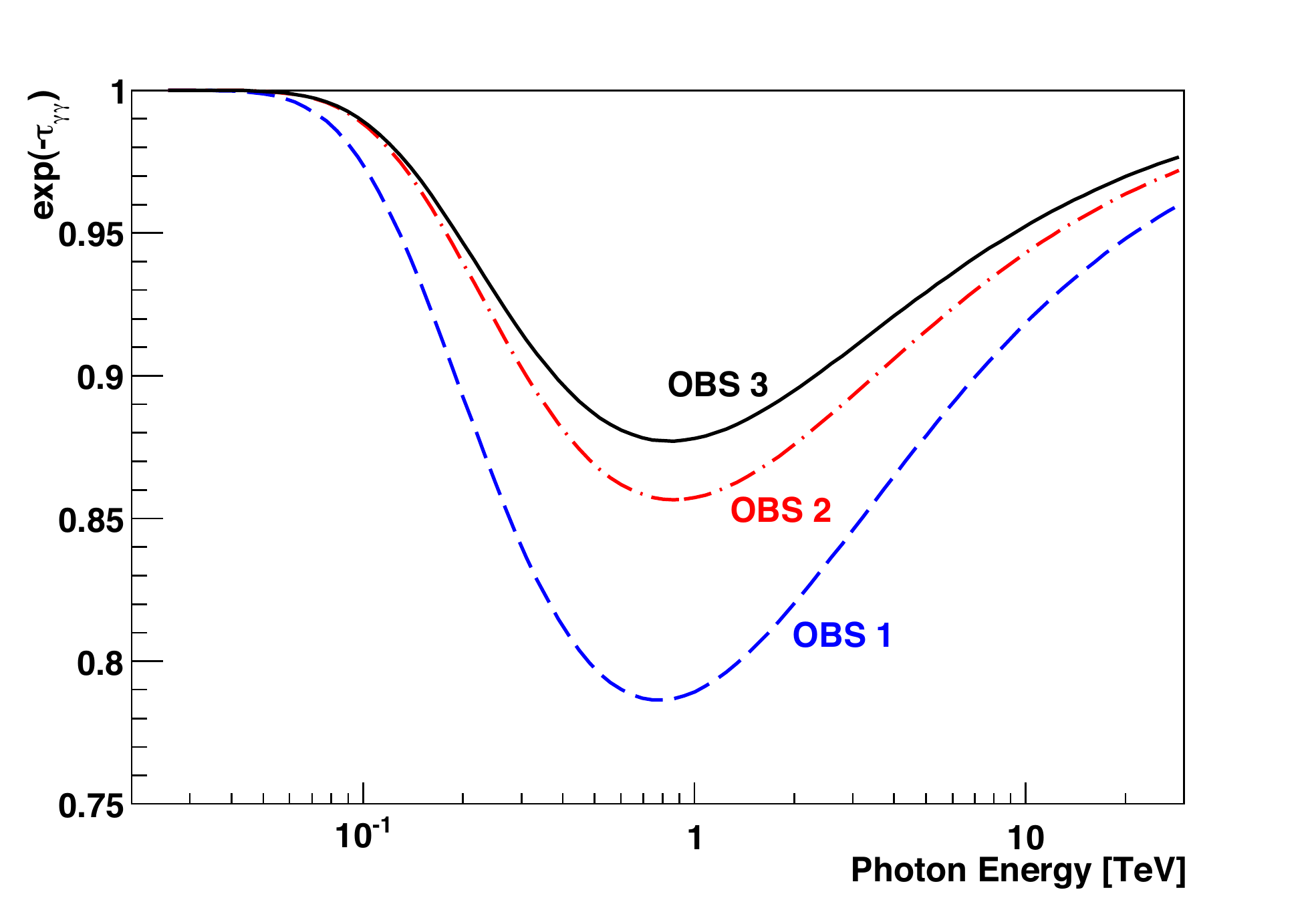}
\includegraphics[width=0.53\textwidth]{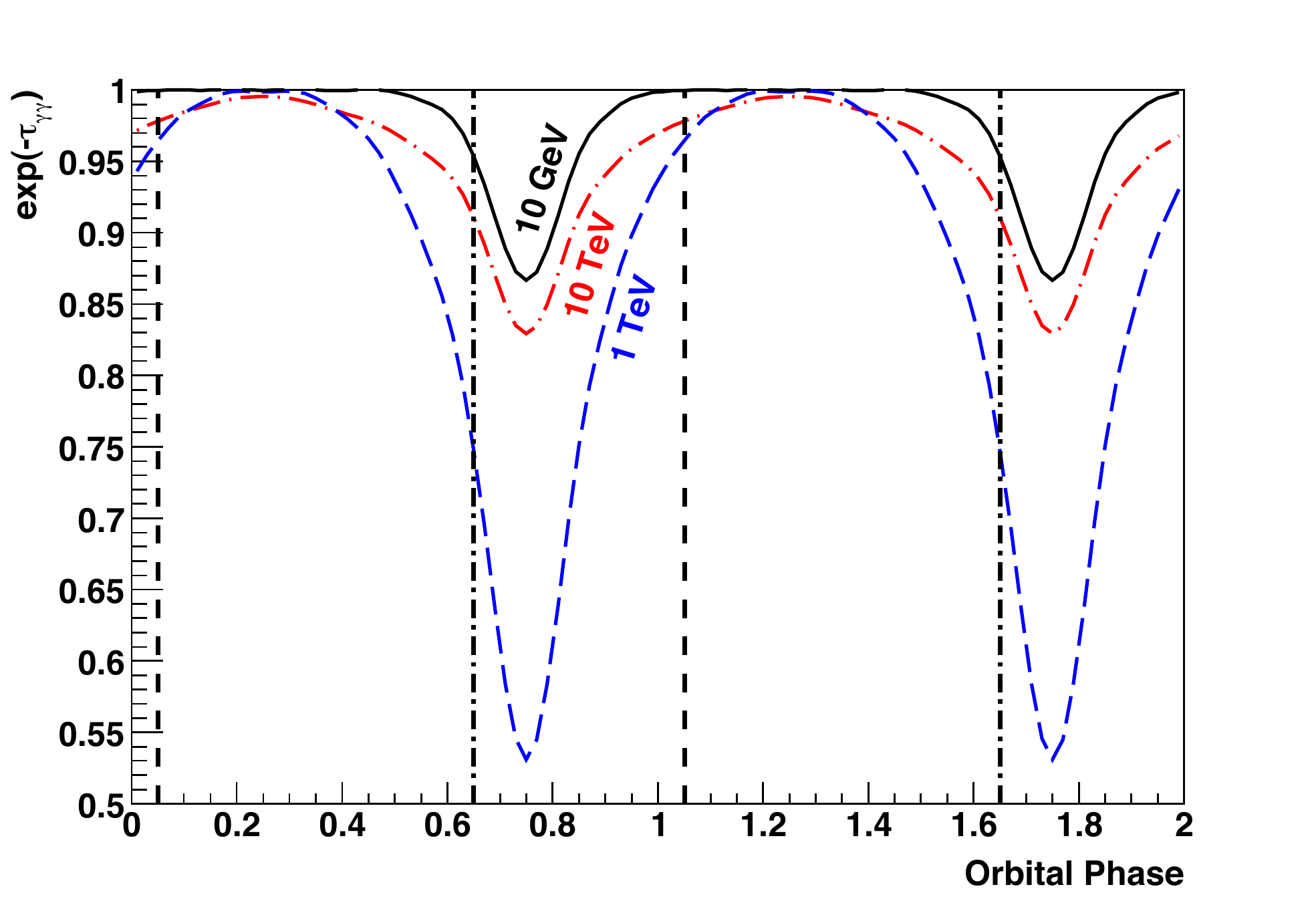}
\caption{Levels of \gr\ absorption due to pair production with stellar photons as predicted by the numerical model outlined by \cite{2006A&A...451....9D}. \textit{Top panel:} expected \gr\ transmission as a function of photon energy for Cir X-1 assuming an inclination $i=66^{\circ}$ and using the best-fit ephemeris derived by \cite{2007MNRAS.374..999J}, which is appropriate for a mid-B supergiant companion. The individual curves correspond to the orbital phases of the first ({blue dashed}), second (red dot-dashed), and third (black solid) \hess\ observation intervals. \textit{Bottom panel:} expected \gr\ transmission for V4641 Sgr as a function of orbital phase derived using the orbital solution of \citep{2001ApJ...555..489O} and assuming a circularised orbit. The individual curves represent photon energies of 10 GeV ({black solid}), 1 TeV ({blue dashed}), and 10 TeV ({red dot-dashed}). Vertical lines indicate the first ({dot-dashed}) and second ({dashed}) \hess\ observation epochs.}
\label{fig:mqs:abs}
\end{center}
\end{figure}

Absorption of $\gamma$-rays by pair production is expected to be negligible in GRS 1915+105, since the donor star is too cool and faint to produce a strong ultraviolet photon field. The same is true of Cir X-1 if the conventional assumption of a low-mass companion is adopted. 
For completeness, Figure \ref{fig:mqs:abs} plots the level of \gr\ absorption predicted by a numerical implementation of the model presented by \cite{2006A&A...451....9D}, assuming that the companion star in Cir X-1 is in fact a mid-B supergiant as proposed by \cite{2007MNRAS.374..999J}. The separate curves are representative of the three orbital phase intervals corresponding to the \hess\ observation epochs, and were derived using the system parameters derived by \cite{2007MNRAS.374..999J} in conjunction with typical values for the temperature ($T_{\rm eff}\approx20000$ K) and radius ($R\approx9$ R$_{\odot}$) of a mid-B supergiant. It is evident that some non-negligible absorption is expected, particularly during the first observation interval. Nonetheless, it seems unlikely that the expected levels of attenuation ($\lesssim20\%$) would suppress an otherwise detectable \gr\ flux sufficiently to yield the low significances listed in Table \ref{tab:mqs:hess_sigmas}.

The situation with regard to \gr\ absorption is clearer in the case of V4641 Sgr, since the companion has been spectroscopically identified as a late B- or early A-type star. Using the system parameters derived by \cite{2001ApJ...555..489O} and assuming a circularised orbit, the model presented by \cite{2006A&A...451....9D} was used to predict the expected levels of \gr\ absorption as a function of orbital phase. As illustrated in Figure \ref{fig:mqs:abs} (bottom panel), absorption might have an important effect during the first \hess\ observation interval, although as with Cir X-1 the predicted levels of absorption ($\lesssim25\%$) would not attenuate a bright \gr\ signal so far below the detection threshold. During the second \hess\ observation interval, when X-ray data show marginal indications of source activity, the predicted absorption due to pair production on the stellar radiation field is negligible.
We note however that, as in the case of Cir X-1, the relative inclination of the jets from V4641 Sgr with respect to the accretion disk may be low \citep{2006ESASP.604..201S}  and, therefore, further absorption of  $\sim100$~GeV -- TeV $\gamma$-ray photons could occur via interaction with the disk thermal photon field (see e.g. \citealp{1992A&A...264..127C}). 

It should also be noted that all the confirmed VHE \gr\ binaries lie at distances of $2-4$ kpc.  In contrast, the targets reported here have maximum distances in the range $7-11$ kpc, resulting in flux dilution factors that are greater by one order of mangnitude. Obviously, this has strong implications for the detectability of any emitted \gr\ signal.

\section{Conclusions}\label{sec:conclusion}

Contemporaneous VHE \gr\ and X-ray observations of GRS 1915+105, Cir X-1, and V4641 Sgr were obtained using \hess\ and \textit{RXTE}. Analysis of the resultant \hess\ data did not yield a significant detection for any of the target microquasars. However, X-ray binaries are dynamic systems and as such are likely to exhibit evolution of their radiative properties, both as a function of orbital phase and also in response to non-deterministic properties. It follows that the non-detections presented in this work do not indicate that the target binary systems do not emit detectable VHE \gr\ emission at phases other than those corresponding to the \hess\ observations.

GRS 1915+105 appears to have been observed during an extended plateau state, the archival multiwavelength data suggesting the presence of continuous, mildly relativistic radio jets at the time of observation. The \textit{RXTE} observations of Cir X-1 yield data that are consistent with strongly varying obscuration of the X-ray source shortly after periastron passage, but these data are not indicative of bright flaring during the \hess\ observation epochs. Conversely, V4641 Sgr appears to have been observed during an episode of mild, transient flaring, although rapid source variability, combined with the limited duration of the strictly simultaneous \hess\ and \textit{RXTE} exposure, complicates interpretation. 

Microquasars continue to be classified as targets of opportunity for IACTs, requiring a rapid response to any external trigger to maximise the likelihood of obtaining a significant detection. These conditions are realised with the commissioning of the H.E.S.S. 28 m telescope, which aims to lower the energy threshold from 100 GeV to about 30 GeV (\cite{Parsons2015}; \cite{Holler2015}; \cite{HollerCrab2015}) while simultaneously enabling very rapid follow-up observations \citep{Hofverberg2013a}. To exploit these new opportunities and an increasing understanding of the behaviour of microquasars, the triggering strategies for TeV follow-up observations have evolved significantly in recent years. In the future, alternative observational strategies, including continuous monitoring of candidate microquasars in the VHE \gr\ band, may become possible using dedicated sub-arrays of the forthcoming Cherenkov Telescope Array \citep{2010arXiv1008.3703C}. 

Irrespective of the non-detections presented herein, the tantalising observations of Cygnus X-3 at GeV energies and Cygnus X-1 by the MAGIC telescope ensures that the motivations for observing microquasars using IACTs remain compelling. Indeed, by further constraining the \gr\ emission properties of microquasars, subsequent observations will inevitably yield an enhanced understanding of astrophysical jet production on all physical scales. More optimistically, the detection of additional \gr -bright microquasars would greatly facilitate a comprehensive characterisation of the particle acceleration and radiative emission mechanisms that operate in such systems. 

\begin{acknowledgements}
The support of the Namibian authorities and of the University of Namibia
in facilitating the construction and operation of H.E.S.S. is gratefully 
acknowledged, as is the support by the German Ministry for Education and
Research (BMBF), the Max Planck Society, the German Research Foundation (DFG), 
the French Ministry for Research, 
the CNRS-IN2P3 and the Astroparticle Interdisciplinary Programme of the 
CNRS, the U.K. Science and Technology Facilities Council (STFC), 
the IPNP of the Charles University, the Czech Science Foundation, the Polish 
Ministry of Science and Higher Education, the South African Department of 
Science and Technology and National Research Foundation, and by the 
University of Namibia. We appreciate the excellent work of the technical 
support staff in Berlin, Durham, Hamburg, Heidelberg, Palaiseau, Paris, 
Saclay, and in Namibia in the construction and operation of the 
equipment.\end{acknowledgements}

\bibliographystyle{aa}
\bibliography{UpperLims}

\appendix
\section{Modelling and determination of the system X-ray states}

\subsection{GRS 1915+105}~\label{app:GRS1915state}

\begin{figure*}
\centering
\includegraphics[width=0.6\textwidth]{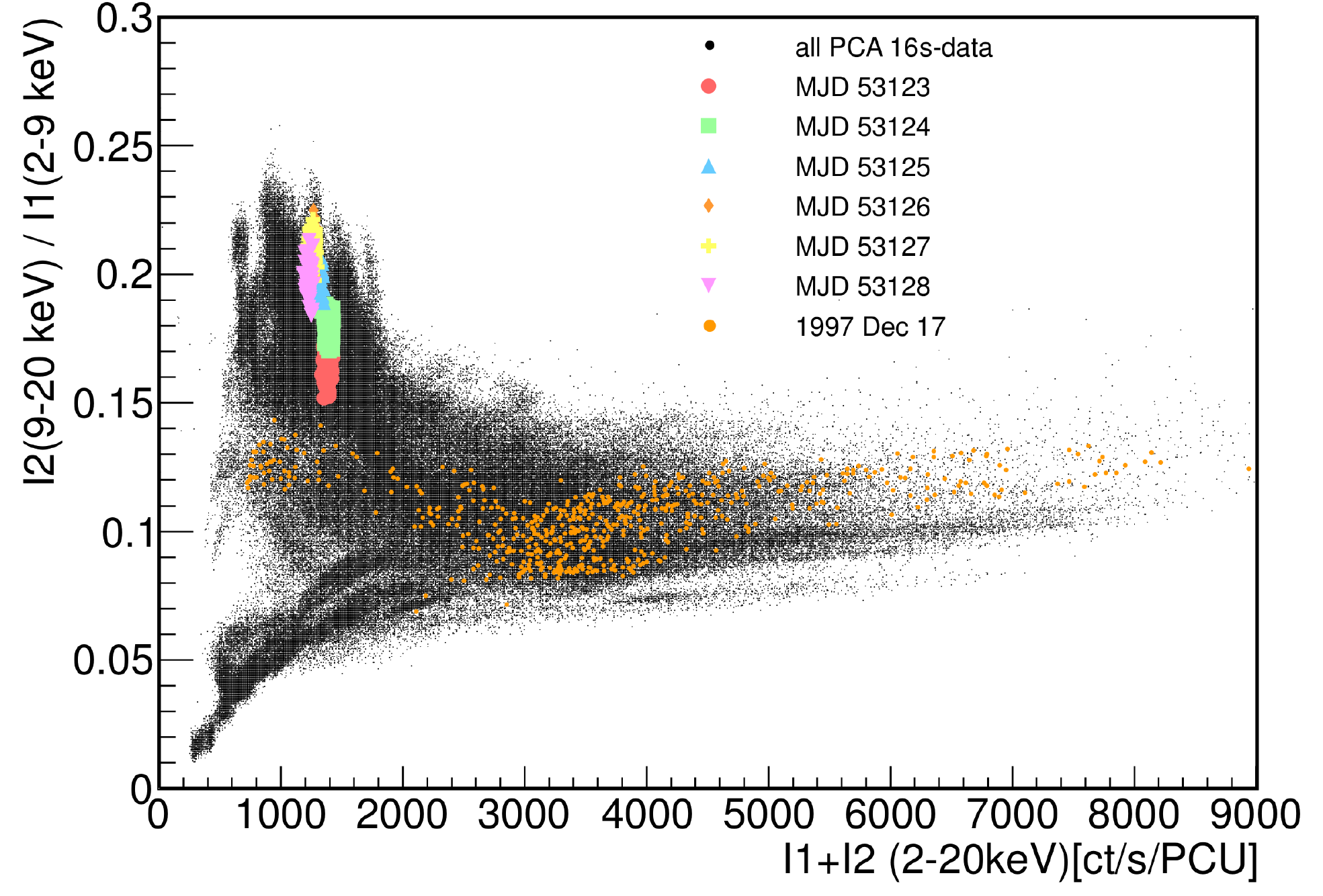}
\caption{Hardness-intensity diagram of GRS 1915+105 derived from the entire archival \textit{RXTE} PCA data set. The temporal sampling is 16 seconds. Hardness is defined as the ratio of fluxes measured in the 2-9 keV and 9-20 keV bands, while intensity is defined as the sum of the band-limited fluxes in units of counts per second. Data corresponding to H.E.S.S. observations are highlighted using symbols to identify the individual days of observation. For comparison, the data corresponding to noteworthy events such as known flares observed by \textit{RXTE} are also plotted (see e.g. \citep{Soleri2006}).}
\label{HID_GRS1915}
\end{figure*}

Figure \ref{HID_GRS1915} shows the HID derived from the entire archival \textit{RXTE} PCA data set for GRS 1915+105. Hardness is defined as the ratio of fluxes measured in the 2-9 keV and 9-20 keV bands, while intensity is defined as the sum of the band-limited fluxes in units of counts per second. The background has been subtracted and the light curves are sampled in 16 s intervals. These definitions are used consistently for the three HIDs presented in this paper.
In the case of GRS 1915+105, the \hess\ observation took place in a low-hard state (LHS; symbols for MJD 53123-8 in top-left sector of Figure \ref{HID_GRS1915}), in which compact jets are expected to be present and characterised by a potentially radio-loud $\chi$ X-ray variability class (belonging to state C following the classification of \citet[][]{2000A&A...355..271B}). For comparison, the orange points in Figure \ref{HID_GRS1915} correspond with data obtained on 17 December  1997. These data were studied by \citet{Soleri2006} who associated them with a hard-intermediate state (HIMS) to soft-intermediate state (SIMS) transition. 

\begin{figure*}
\includegraphics[width=\textwidth]{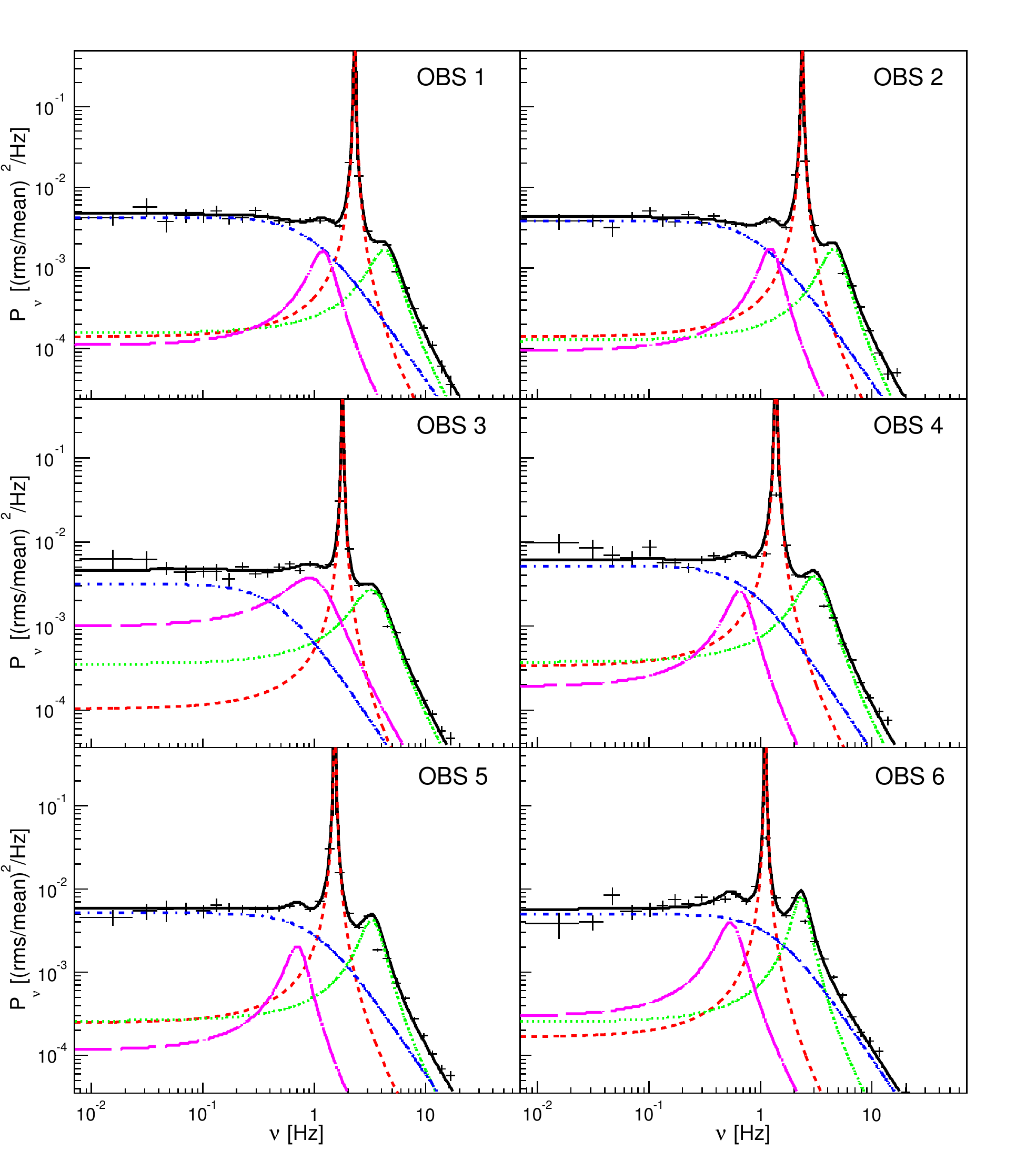}
\caption{X-ray power density spectra (PDS) corresponding to the six \textit{RXTE} observations of GRS 1915+105. The PDS were fitted using a superposition of Lorentzian functions comprising two broad continuum components (blue dot-dashed and green dotted curves) and several narrow QPO peaks (remaining curves). For all six observations, the derived properties of higher frequency continuum component are consistent with the radio-loud $\chi$ state.}\label{fig:grs_powspecs}
\end{figure*}

The power density spectra shown in Figure \ref{fig:grs_powspecs} show the presence of low frequency QPOs: following the approach of \cite{2002ApJ...572..392B}, a Lorentzian decomposition of the observed power spectra was performed using two broad continuum components and several narrower QPO peaks. Crucially, the characteristic frequency ($\nu_{\max} = \sqrt{\nu_{0}^{2}+\Delta^{2}}$ \citep[see][]{2002ApJ...572..392B}) of the higher frequency continuum component never exceeds $\sim 4$ Hz during our observations. This is far below the characteristic cut-off frequencies associated with previous observations of the radio-quiet $\chi$ state \citep[e.g.][]{2001ApJ...558..276T}.
Radio-quiet observations \citep[][]{1997ApJ...488L.109B, 2001ApJ...558..276T} exhibit significant band-limited white noise extending to high frequencies $f\sim60-80$ Hz, while in radio-loud case such noise is either absent or exhibits an exponential cut-off at $\sim15$ Hz \citep[][]{2001ApJ...558..276T}. Consequently, the absence of band limited noise at high frequencies is consistent with a radio-loud state C. 

\begin{figure*}
\includegraphics[width=\textwidth,angle=0,clip]{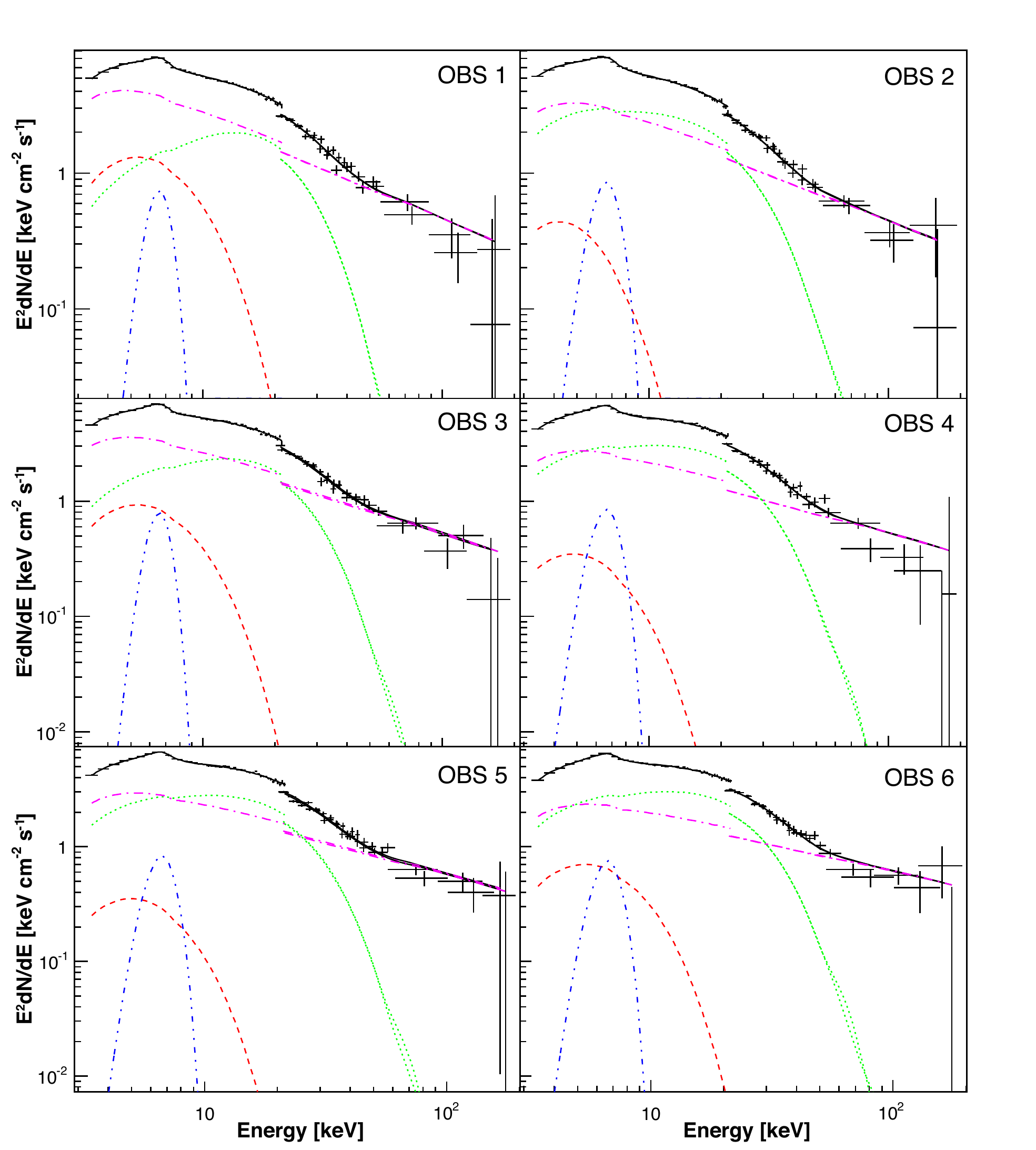}
\caption{Unfolded \textit{RXTE} $3-200\,$keV X-ray spectra of GRS 1915+105 showing the individual \texttt{XSPEC} model components: {\sc DiskBB} (red dashes), {\sc CompST} (green dots), {\sc Powerlaw} (magenta dot-dashed), {\sc Gaussian} (blue double-dot-dashed),  and the total spectrum (black).}\label{fig:grs1915_spec}
\end{figure*}

Figure \ref{fig:grs1915_spec} illustrates the spectral analysis performed for each GRS 1915+105 \textit{RXTE} observation. The simple model described by \cite{2001A&A...372..793V} has been adopted: a continuum model comprising a disk black-body ({\sc DiskBB}\footnote{\url{http://heasarc.gsfc.nasa.gov/xanadu/xspec/manual/XspecModels.html}}) component, a hybrid thermal and a non-thermal Comptonisation component ({\sc CompST}), and a separate power law ({\sc Powerlaw}) to model the high-energy emission. Interstellar absorption was modelled using the {\sc Wabs} model in \texttt{XSPEC} with the equivalent hydrogen column fixed to a value of $6\times10^{22}$ cm$^{-2}$ \citep{1997ApJ...488L.109B,1999ApJ...513L..37M,1999ApJ...527..321M}. A constant multiplicative factor was introduced to account for the normalisation of HEXTE relative to the PCA.
The addition of a {\sc Gaussian} component with centroid energy $E_{\rm Line}$ fixed at 6.4 keV was found to significantly improve the resultant model fit. 

As demonstrated by the reduced $\chi^{2}$ values listed in Table \ref{tab:grs1915_free_pars}, the fitted model provides an adequate description of the \textit{RXTE} data, and the derived model parameters correspond closely to those  obtained by \cite{2001A&A...372..793V} during earlier episodes of radio-loud $\chi$-state behaviour, supporting the attribution of this state to the epochs of \hess\ observation.
However, insufficient event statistics prevent the inference of robust conclusions regarding the origin of the X-rays that were observed in this study. A more sophisticated and physically well-motivated model, describing an X-ray corona ({\sc EQPAIR} from \cite{2000HEAD....5.2311C}) gives a similar goodness-of-fit, after accounting for a larger parameter count.
Physically, the power-law component could result from Comptonisation by energetic electrons in a corona or could be generated by synchrotron radiation at the base of a jet.
The former scenario is discussed in a number of papers (e.g. \cite{Zdziarski2003}), while the latter was studied  by \cite{IAU:8143332} in the context of a similar plateau state of GRS1915+105. They applied a leptonic jet model \citep{2005ApJ...635.1203M} to X-ray, IR, and radio data.
Although their model provided statistically convincing broadband fits, this was only achievable when adopting extreme parameter values.
The power law with $\Gamma\simeq 2.7$ that was derived in this study (Table \ref{tab:grs1915_free_pars}) cannot be extrapolated down to UV band without generating an inconsistency in subsequently inferred bolometric luminosity. It should be interpreted as a phenomenological approximation of a high-energy tail, which itself might only be partially accounted for by SSC radiation.

A similar plateau state of GRS 1915+105 (October 1997) was studied by \cite{2002MNRAS.331..745K}, who attributed the observed radio emission to quasi-continuous ejecta forming the compact jet.

\begin{table*}[thp!]
\caption{\label{tab:GRS_pars}\texttt{XSPEC} model components and best-fit parameters for GRS 1915+105. As discussed in the text, an additional {\sc Wabs} component (with equivalent hydrogen column density fixed to $N_{\rm H} = 6\times10^{22}$cm$^{-2}$) was used to model the effects of interstellar absorption. The parameter errors correspond to a $\Delta\chi^{2}$ of 2.71. Frozen parameters are indicated by *.}\label{tab:grs1915_free_pars}
\centering
\begin{tabular}{lllll}
\hline\hline
{\sc Component} &       {\sc Parameter} & {\sc OBS 1} & {\sc OBS 2} &  {\sc OBS 3} \\
\hline
{\sc DiskBB}  &          $T_{\rm{in}}$  [keV] &      ${1.695}^{+0.42}_{-0.59}$  &       ${1.137}^{+0.6}_{-0.3}$ &       ${1.698}^{+0.63}_{-0.77}$       \\
{\sc DiskBB}  &         Norm  &      ${31.1}^{+120}_{-17}$      &       ${81.3}^{+360}_{-78}$   &       ${20.9}^{+200}_{-16}$   \\
{\sc CompST}  &          $kT_{e}$ [keV]  &       ${4.195}^{+2}_{-0.54}$ &       ${5.057}^{+1.1}_{-0.81}$        &       ${4.460}^{+0.54}_{-0.7}$\\
{\sc CompST}  &          $\tau$  &       ${13.145}_{-6.9}$      &       ${7.801}^{+3.7}_{-1.8}$ &       ${10.959}_{-4.7}$\\
{\sc Powerlaw}  &          $\Gamma_{\rm phot}$  &       ${2.714}^{+0.058}_{-0.86}$      &       ${2.668}^{+0.12}_{-0.15}$       & ${2.644}^{+0.077}_{-0.82}$\\
{\sc Gaussian}  &          $E_{\rm Line}$ [keV] &\multicolumn{3}{c}{\rule[0.5ex]{0.135\textwidth}{0.5pt} 6.4* \rule[0.5ex]{0.135\textwidth}{0.5pt}}\\
{\sc Gaussian}        &   $\sigma$ [keV]       &       ${0.776}^{+0.24}_{-0.29}$        &       ${0.891}^{+0.15}_{-0.17}$ &       ${0.730}^{+0.22}_{-0.24}$\\
{\sc Gaussian}      &          W [eV] &       66.44  &       87.76  &       70.44\\\hline
$\chi^{2}_{\nu}$ (NDF) &                &       0.82 (77) & 0.96 (78) & 0.99 (80)\\
\hline\hline
{\sc Component} &       {\sc Parameter} & {\sc OBS 4} & {\sc OBS 5} &  {\sc OBS 6} \\
\hline
{\sc DiskBB}  &          $T_{\rm{in}}$  [keV] &       ${1.350}^{+0.63}_{-0.42}$ &       ${1.415}^{+0.63}_{-0.51}$ &       ${1.713}^{+0.51}_{-0.55}$\\
{\sc DiskBB}  &          Norm  &       ${30.8}^{+140}_{-28}$    &       ${23.3}^{+110}_{-23}$ &       ${15.4}^{+28}_{-13}$\\
{\sc CompST}  &          $kT_{e}$ [keV] &      ${5.130}^{+0.77}_{-0.69}$        &       ${4.888}^{+0.73}_{-0.69}$ &       ${5.118}^{+0.83}_{-0.82}$\\
{\sc CompST}  &          $\tau$  &       ${8.516}^{+4.4}_{-1.6}$        &       ${8.870}^{+6.1}_{-1.7}$ &       ${9.028}^{+14}_{-1.9}$\\
{\sc Powerlaw}  &          $\Gamma_{\rm phot}$  &       ${2.503}^{+0.12}_{-0.38}$       &       ${2.521}^{+0.026}_{-0.27}$      &       ${2.441}^{+0.13}_{-0.37}$\\
{\sc Gaussian}  &          $E_{\rm Line}$ [keV] &\multicolumn{3}{c}{\rule[0.5ex]{0.135\textwidth}{0.5pt} 6.4* \rule[0.5ex]{0.135\textwidth}{0.5pt}}\\
{\sc Gaussian}        &   $\sigma$ [keV]       &       ${0.894}^{+0.16}_{-0.19}$        &       ${0.902}^{+0.16}_{-0.18}$       &       ${0.928}^{+0.19}_{-0.24}$\\
{\sc Gaussian}      &          W [eV] &       91.15  &       91.78  &       86.93\\\hline
$\chi^{2}_{\nu}$ (NDF) &                &      1.12 (80) & 1.19 (83) & 0.82 (80)\\\hline

\end{tabular}
\end{table*}

\begin{figure*}
\centering
\includegraphics[width=0.7\textwidth]{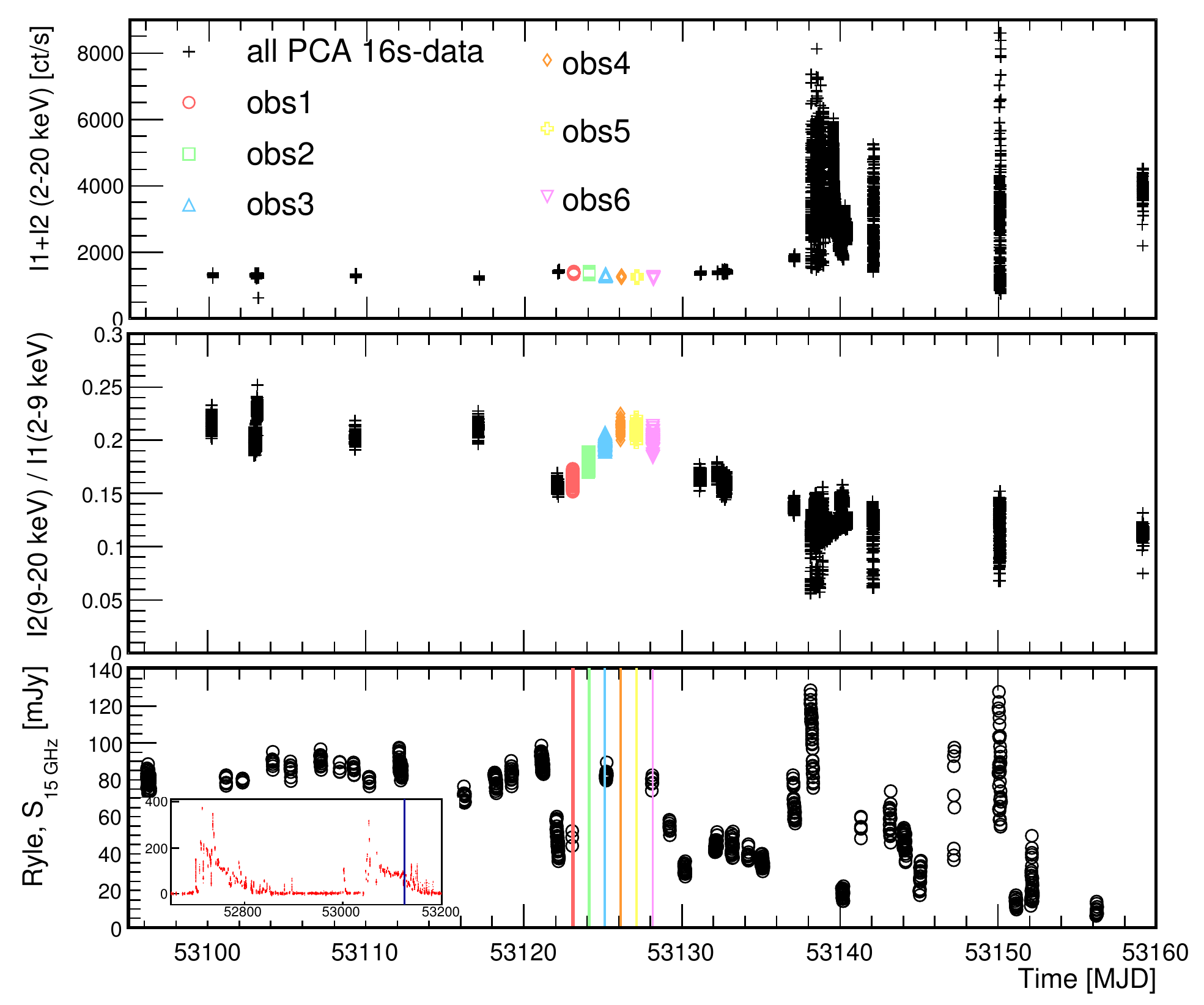}
\caption{Evolution of GRS 1915+105 before and after H.E.S.S. observations. \textit{RXTE} PCA intensity and hardness (defined in the caption of Figure \ref{HID_GRS1915}) are plotted on the top and middle panels, respectively. The bottom panel presents contemporaneous 15 GHz radio data obtained using the Ryle Telescope \citep{Pooley2006}. Long-term monitoring data in the 15 GHz radio band are illustrated in the inset. Coloured markers and lines indicate the H.E.S.S. observation epochs.}
\label{evol_GRS1915}
\end{figure*}

In Figure \ref{evol_GRS1915}, the 15 GHz radio surface brightness and the X-ray hardness and intensity corresponding with the epochs of \hess\ observation are illustrated in a broader historical context \citep{Pooley2006}\footnote{ \url{http://www.mrao.cam.ac.uk/~guy/} }.
The \hess\ observations started approximately one week before the end of a long radio-loud plateau and were triggered by a transient dip in radio flux whilst the plateau end was not yet reached. The plateau ended about two days after the last \hess\ observation, followed by a radio and X-ray flare ten days later.

In summary, GRS 1915+105 remained in a radio-loud $\chi$ state with steady, mildly relativistic jets at the time of \hess\ observations without clear signs of a state transition.

\subsection{Cir X-1}~\label{app:CirState}

\begin{figure*}[thp!]
\centering
\includegraphics[width=0.6\textwidth]{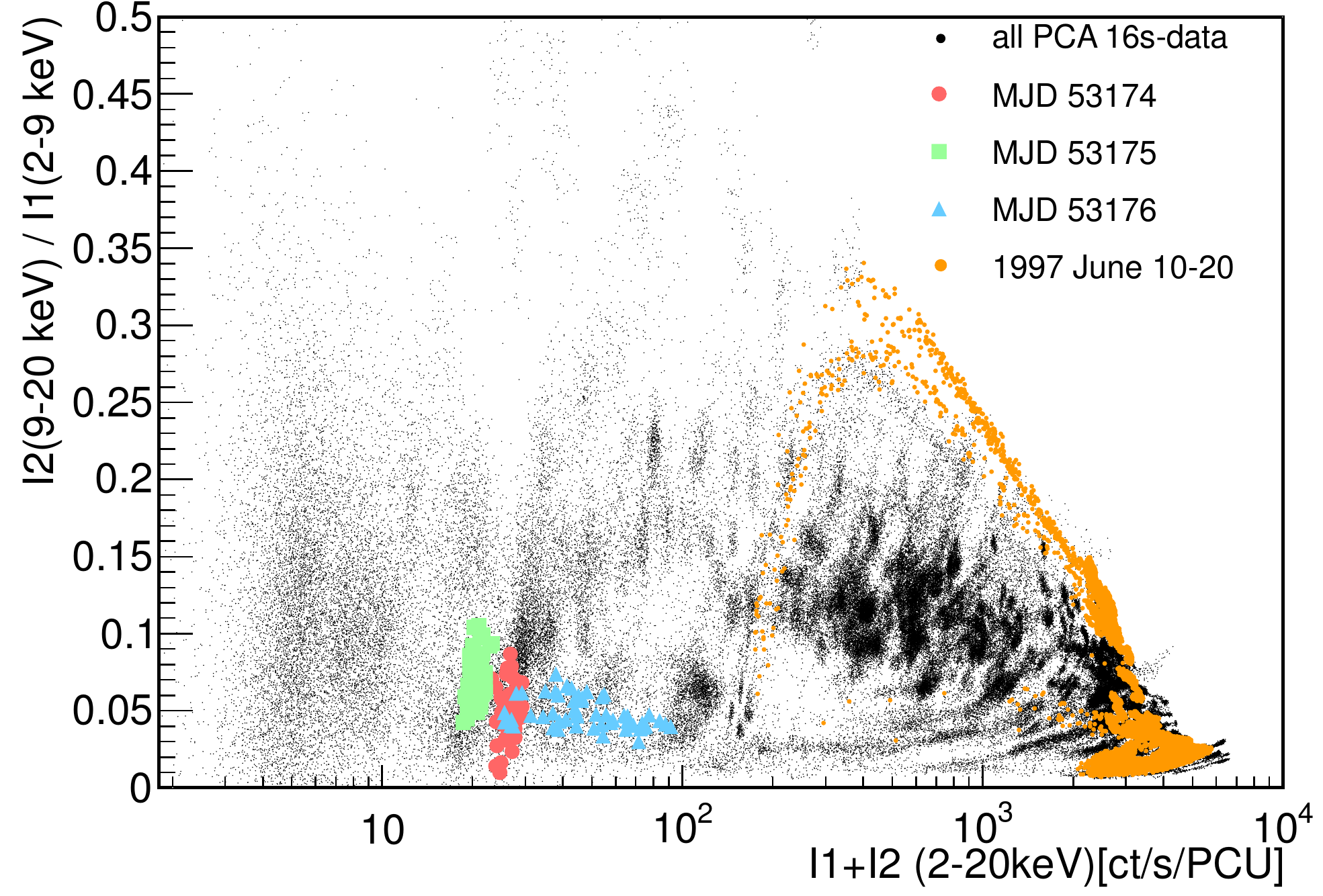}
\caption{Hardness-intensity diagram derived from the entire archival \textit{RXTE} PCA data set for Circinus X-1. Intensity and hardness are defined in the caption of Figure \ref{HID_GRS1915}. Coloured markers indicate data that correspond with the H.E.S.S. observation epochs and historically noteworthy episodes of flaring activity (e.g. \cite{1999ApJ...517..472S}).}
\label{HID_CirX1}
\end{figure*}

The HID for Cir X-1 is shown in Figure \ref{HID_CirX1}. An extensive study, examining ten days out of the 16.55 orbital period was performed by \citet{1999ApJ...517..472S} in 1997; the corresponding \textit{RXTE} data are indicated with orange symbols on figure \ref{HID_CirX1}. The study focused on the toroidally distributed data plotted in the lower right part of the HID. 
\citet{1999ApJ...517..472S}, studied the spectral and temporal X-ray evolution of Cir X-1 along three distinct branches (horizontal, normal, flaring) in the HID. This evolution occurred during a half-day period, approximately one day after periastron and was repeated few days later. Such behaviour is typical of a "Z source".

In \citet{1999ApJ...517..472S}, the periastron passage corresponds to the data at low flux and hardness (dipping episode, lower-left part of the cycle). 
Data contemporaneous to the \hess\ observations (MJD 53174-6) are indicated by symbols and exhibit low X-ray intensity and hardness values. They span two days starting 19 hours after the periastron, which coincides with the orbital range explored by \citet{1999ApJ...517..472S}, but in a much fainter X-ray luminosity context.

\begin{figure*}[thp!]
\includegraphics[width=\textwidth]{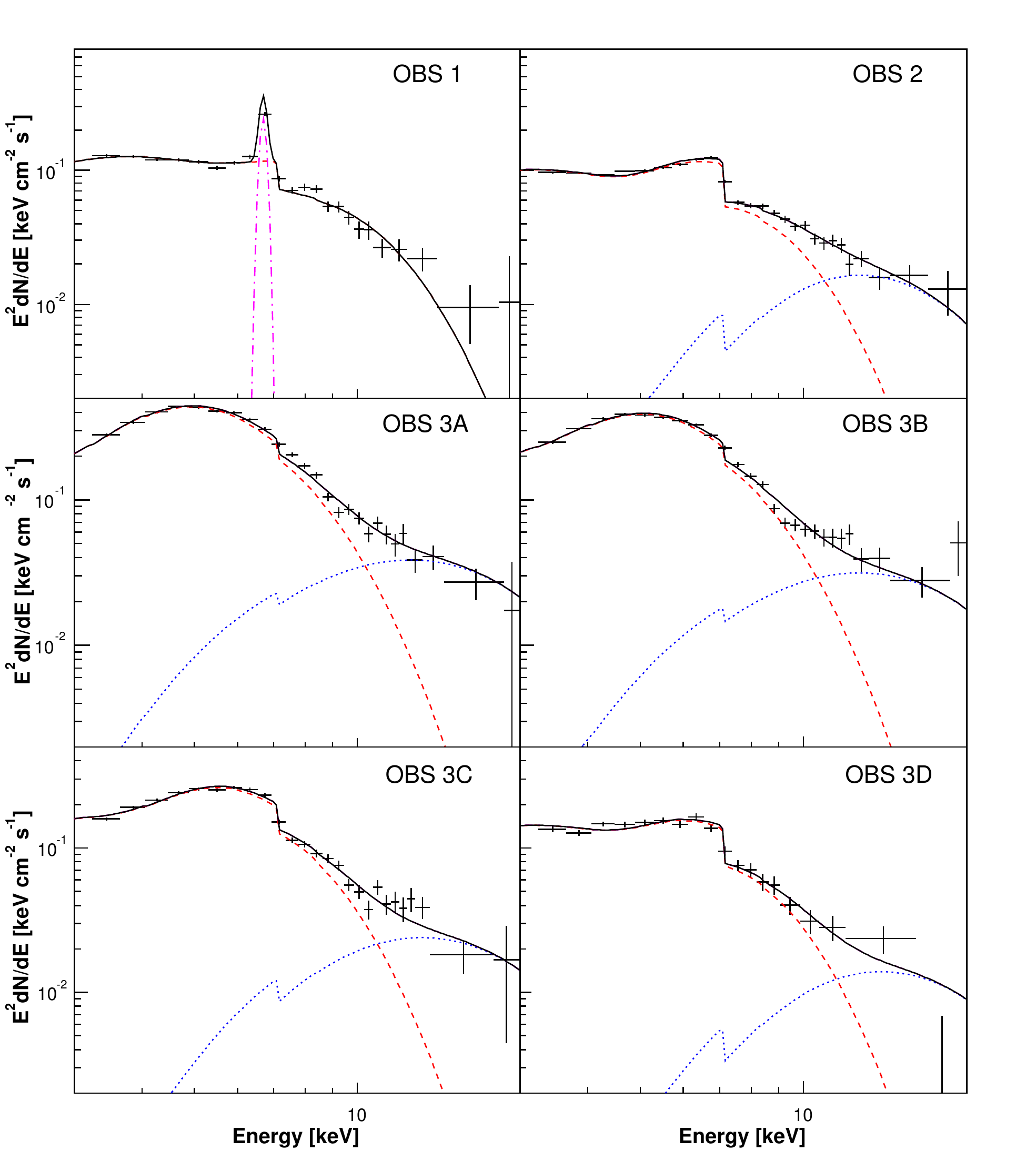}
\caption{Unfolded $3-20\,$keV \textit{RXTE} X-ray spectra of Cir X-1 for the first and second observations (top two panels), and the four sub-intervals of the third observation (bottom four panels). The solid black curves illustrate the total spectral model, while the individual components are represented as follows: {\sc DiskBB} (red dashed), {\sc Bbody} (blue dotted), {\sc Gaussian} (pink dot-dashed lines).}\label{fig:cirx1}
\end{figure*}

Inspection of the $3-20\,$keV PCA spectra shown in Figure \ref{fig:cirx1} reveals that the observed flux variability is accompanied by marked variations in spectral shape. For the third observation, individual spectra were extracted from the four regions (A to D) shown in Figure \ref{fig:cir_lightcurves} (bottom-right panel), segregated on the basis of average $2-20\,$keV count rates. Fitting of the spectral data from the third observation employed a similar approach to that of \citet{1999ApJ...524.1048S} with the unabsorbed continuum modelled using a disk black-body component ({\sc{DiskBB}} in \texttt{XSPEC}) at low energies in combination with a single temperature black body ({\sc{Bbody}}) that dominates above $\sim15$ keV.

Previous observations of Cir X-1 during periastron dips \citep[e.g.][]{1999ApJ...524.1048S, 2008ApJ...672.1091S} reveal the evidence of strong, complex, and variable intrinsic X-ray absorption. Consequently, diagnosis of the system behaviour during the third \textit{RXTE} observation is critically dependent upon whether the observed variability represents a genuine change in the underlying continuum emission or is simply an artefact of varying absorption. Accordingly, two components are used to separately simulate intrinsic and extrinsic X-ray absorption characteristics. The bipartite intrinsic absorption is treated using a partial covering model ({\sc{Pcfabs}}), while a simple photoelectric model ({\sc{Wabs}}) simulates the absorbing effect of the interstellar medium. Adopting a weighted average of the neutral hydrogen data of \citet{2005A&A...440..775K} calculated using the \texttt{nH} ftool, a fixed effective hydrogen column with $N_{H}=1.59\times10^{22}~{\rm cm}^{-2}$ was assumed for the {\sc{Wabs}} component. This assumption is consistent with an estimate of the surrounding interstellar medium density used by \citet{2006MNRAS.372..417T} to model the evolution of the radio nebula of Cir X-1. 
 
In order to constrain the origin of the observed spectral variability, a joint fit was performed using the complete best-fitting model. Initially, the continuum and extrinsic absorption components ({\sc{DiskBB}}, {\sc{Bbody}}, {\sc{Wabs}}) were constrained to be equal for all individual spectra, while the component related to intrinsic absorption ({\sc{Pcfabs}}) was allowed to vary independently. Although this model provides a reasonable fit to the observational data ($\chi^{2}_{\nu} = 1.38$), allowing the normalisation of the {\sc{Bbody}} component to vary between observations improves the fit quality somewhat, yielding $\chi^{2}_{\nu} = 1.27$. The parameters that result from fitting this more relaxed model are listed in Table \ref{tab:cir_x1_pars_obs3}. A statistical comparison of the alternative model fits using the F test yields a $\sim1\%$ probability that the observed improvement in fit quality would be obtained even if the more restrictive model was correct. This marginal evidence for variation of the {\sc{Bbody}} component normalisation might indicate rapid fluctuations of the X-ray continuum above $\sim 10$ keV. However, the available data cannot exclude an alternative scenario in which apparent changes in the fitted {\sc{Bbody}} parameters arise purely from imperfect modelling of substantial variations in the intrinsic X-ray absorption with no requirement for genuine evolution of the underlying continuum.

Table \ref{tab:cir_x1_pars} lists the parameters of the spectral fits obtained from the first and second observations. A similar continuum model to that obtained from the third observation also provides a good fit ($\chi^{2}_{\nu} = 1.22$) to the spectrum obtained during the second observation. In contrast, the spectrum obtained during the first observation is more appropriately described by a single, heavily absorbed disk black-body component with large correlated residuals around $\sim6.5$ keV statistically favouring the addition of a {\sc{Gaussian}} line component. This continuum variability is consistent with the results of \citet{1999ApJ...517..472S} who found that significant variation of the continuum parameters could occur on timescales of a few hours.

Overall, the \textit{RXTE} data reinforce the accepted paradigm of enhanced mass transfer during the periastron passage of the compact primary with the strong and variable intrinsic absorption attributed to obscuration by a turbulent accretion flow \citep[see e.g.][]{ooster,1980A&A....87..292M,2001ApJ...561..321I}. A marked disparity between best-fitting model components and parameters of the first and second observations implies a dramatic evolution of the local radiative environment. A $\sim30\%$ decrease in continuum luminosity accompanied by a similar reduction of the intrinsic absorption column suggests a significant decrease in the mass transfer rate. Subsequent fluctuation in the inferred magnitude of the absorption column during the third observation is indicative of dispersion or reorganisation of the recently accreted material.

\newpage\clearpage

\newpage\clearpage

\begin{table*}[th!]
\centering
\caption{\label{tab:cir_x1_pars_obs3} Spectral parameters for Cir X-1 during OBS 3 corresponding to the orbital phase interval $0.1718\leq\phi\leq0.1725$ (according to the ephemeris of \cite{2007ATel..985....1N}). \texttt{XSPEC} model components, best-fit parameters, and $3-20\,$keV model fluxes are shown for the four separate sub-intervals illustrated in Figure \ref{fig:cir_lightcurves} in order of decreasing model flux. As discussed in the text, an additional {\sc Wabs} component (with equivalent hydrogen column density fixed to $N_{\rm H} = 1.59\times10^{22}$cm$^{-2}$) was used to model the effects of interstellar absorption. Jointly fitted parameters assume the values quoted for OBS 3A. The parameter errors correspond to a $\Delta\chi^{2}$ of 2.71.}
\renewcommand{\arraystretch}{1.14}
\begin{tabular}{llllll}
\hline\hline
{\sc Component}                 &       {\sc Parameter}                                 & OBS 3A                                                  & OBS 3B                                                        & OBS 3C                                                  & OBS 3D \\
\hline
{\sc DiskBB}                    &        $T_{\rm in}$ [keV]                             &$1.059_{-0.04}^{+0.03}$                                         &\multicolumn{3}{c}{\multirow{2}{*}{\rule[0.5ex]{0.18\textwidth}{0.5pt} Joint fit \rule[0.5ex]{0.18\textwidth}{0.5pt}}} \\
                                        &        Norm                                           &$(1.914_{-0.44}^{+0.63})\times10^{2}$                 &\\
\hline
{\sc Bbody}                     &        $kT$ [keV]                                     &       $2.954_{-0.47}^{+0.67}$                                 &\multicolumn{3}{c}{\multirow{1}{*}{\rule[0.5ex]{0.18\textwidth}{0.5pt} Joint fit \rule[0.5ex]{0.18\textwidth}{0.5pt}}}\\                               
                                        &        Norm                                           &       $(1.126_{-0.18}^{+0.24})\times10^{-3}$  & $(9.373_{-1.70}^{+2.09})\times10^{-4}$ & $(7.646_{-1.62}^{+2.01})\times10^{-4}$ & $(4.992_{-2.44}^{+2.61})\times10^{-4}$\\      
{\sc Pcfabs}                    &        $N_{\rm H}$ [$\times10^{22}$]  &       $(2.303_{-0.34}^{+0.33})\times10^{1}$   &       $(2.880_{-0.32}^{+0.32})\times10^{1}$   &       $(4.606_{-0.29}^{+0.29})\times10^{1}$   &       $(7.998_{-0.48}^{+0.51})\times10^{1}$\\ 
                                        &        CvrFract                                               &       $(8.630_{-0.20}^{+0.22})\times10^{-1}$                          &       $(8.411_{-0.20}^{+0.19})\times10^{-1}$  &       $(8.726_{-0.16}^{+0.15})\times10^{-1}$  &       $(8.852_{-0.15}^{+0.14})\times10^{-1}$\\        
\hline
$\chi^{2}_{\nu}$ (NDF)  &                                                                       &  1.27 (145)                                                     &\multicolumn{3}{c}{\rule[0.5ex]{0.18\textwidth}{0.5pt} Joint fit \rule[0.5ex]{0.18\textwidth}{0.5pt}} \\
\hline
Model flux                      &[erg cm$^{-2}$s$^{-1}$]                        & $6.121\times10^{-10}$                                   & $5.512\times10^{-10}$                         & $3.822\times10^{-10}$                           & $2.515\times10^{-10}$ \\
\hline
\end{tabular}
\end{table*}

\begin{table*}[]
\centering
\caption{\label{tab:cir_x1_pars} \texttt{XSPEC} model components, best-fit parameters, and $3-20\,$keV model fluxes for Cir X-1 during OBS 1 and OBS 2 corresponding to the orbital phase intervals $0.0486\le\phi\le0.0498$ and $0.1104\le\phi\le0.1112,$ respectively (according to the ephemeris of \cite{2007ATel..985....1N}). As discussed in the text, an additional {\sc Wabs} component (with equivalent hydrogen column density fixed to $N_{\rm H} = 1.59\times10^{22}$cm$^{-2}$) was used to model the effects of interstellar absorption. Parameters marked by * are fixed to the best-fitting values from the third observation (See Table \ref{tab:cir_x1_pars_obs3}). The value of the {\sc Gaussian} $\sigma$ parameter (marked by a $\dagger$ symbol) was also fixed. The parameter errors correspond to a $\Delta\chi^{2}$ of 2.71.}
\renewcommand{\arraystretch}{1.14}
\begin{tabular}{llllll}
\hline\hline
{\sc Component}                 &       {\sc Parameter}                                 & OBS 1                                           & OBS 2 \\
                                        &                                                               &($0.0486\le\phi\le0.0498$)                             & ($0.1104\le\phi\le0.1112$) \\
\hline
{\sc DiskBB}                    &        $T_{\rm in}$ [keV]                             &       $1.355_{-0.08}^{+0.18}$                         &       $1.059$*        \\
                                        &        Norm                                           &       $(3.798_{-3.64}^{+3.42})\times10^{1}$   &       $(1.914)\times10^{12}$* \\
\hline
{\sc Bbody}                     &        $kT$ [keV]                                     &       -                                                               &       $2.465_{-0.39}^{+0.50}$         \\              
                                        &        Norm                                           &       -                                                               &       $(7.577_{-1.22}^{+1.90})\times10^{-4}$\\
\hline
{\sc Pcfabs}                    &        $N_{\rm H}$ [$\times10^{22}$]  &       $(1.353_{-0.22}^{+0.44})\times10^{2}$   & $(9.545_{-0.22}^{+0.25})\times10^{1}$\\
                                        &        CvrFract                                               &       $(8.292_{-2.01}^{+0.87})\times10^{-1}$  & $(9.191_{-0.01}^{+0.01})\times10^{-1}$\\
\hline
{\sc Gaussian}                  &        $E_{\rm Line}$ [keV]                   &       $6.696_{-0.08}^{+0.09}$                         &       - \\
                                        &        $\sigma$ [keV]                                 &       $0.1^{\dagger}$                         &       - \\
                                        &        Norm                                           &       $(1.435_{-0.33}^{+0.36})\times10^{-3}$  &       - \\
\hline
$\chi^{2}_{\nu}$ (NDF)  &                                                                       & 1.07 (34)                                                       & 1.22 (36) \\
\hline
Model flux                      &[erg cm$^{-2}$s$^{-1}$]                        &       $2.722\times10^{-10}$                           & $1.912\times10^{-10}$ \\
\hline
\end{tabular}
\end{table*}

\newpage\clearpage

\begin{figure*}
\centering
\includegraphics[width=0.49\textwidth]{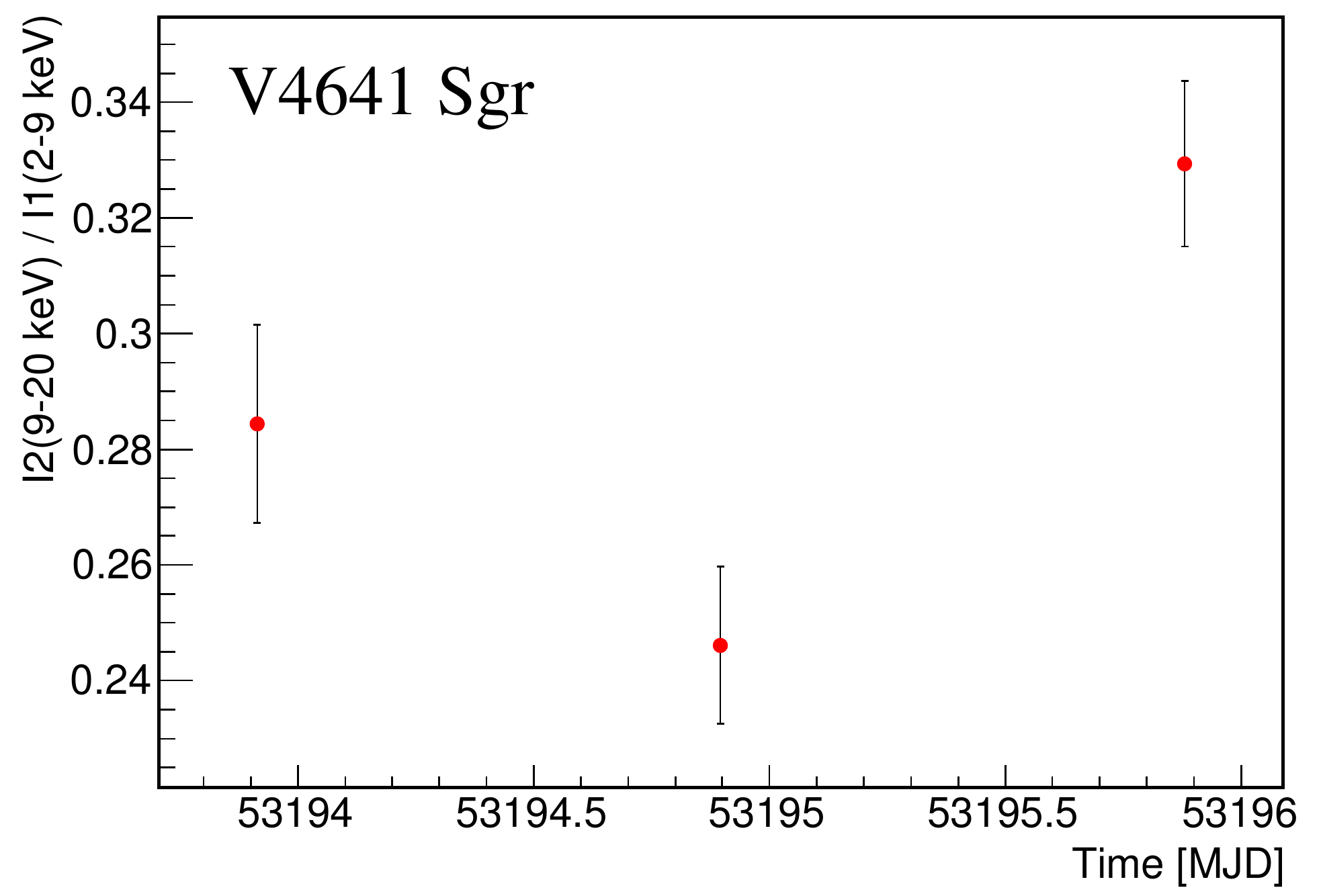}
\includegraphics[width=0.49\textwidth]{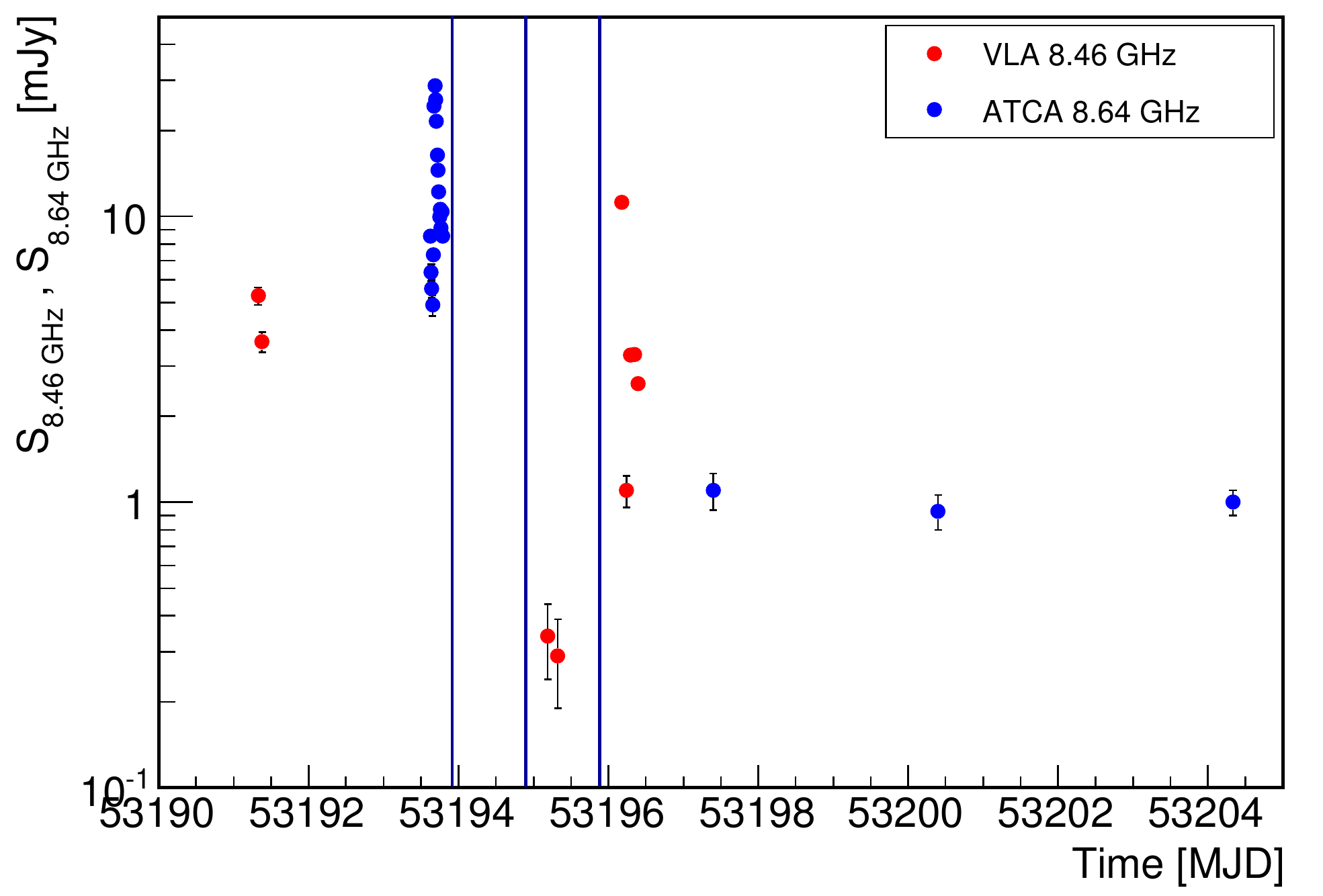}
\caption{{\bf Left:} hardness (ratio of bands 9-20 keV / 2-9 keV ) computed from \textit{RXTE}-PCA data for the three observations of V4641 Sgr. Each point is the mean value of hardness and the error bar represents the standard error on the mean. {\bf Right:} radio data for V4641 Sgr, from ATCA and VLA \citep{2004ATel..296....1R, 2004ATel..302....1S,2004ATel..303....1R}. \textit{RXTE}-PCA and \hess\ observation times are indicated by blue shaded bands.}
\label{fig:V4641hardness_radio}
\end{figure*}
\begin{figure*}
\centering
\includegraphics[width=0.6\textwidth]{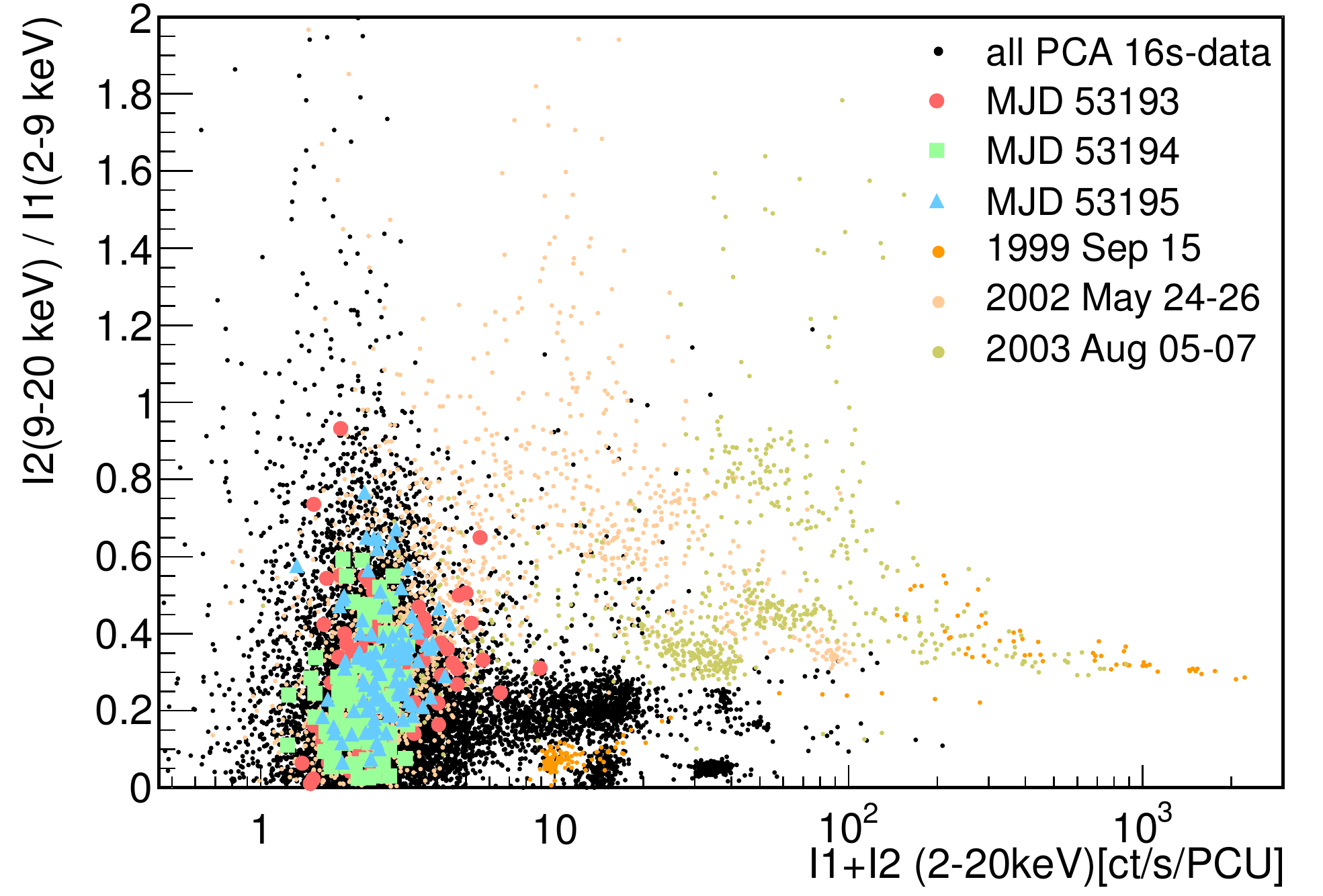}
\caption{Hardness-Intensity diagram derived from the entire archival \textit{RXTE} PCA data set for V4641 Sgr. Intensity and Hardness are defined in the caption of Figure~\ref{HID_GRS1915}. Coloured markers indicate data that correspond with the H.E.S.S. observation epochs, as well as historically noteworthy episodes of flaring activity (e.g. \cite{Wijnands2000}).}
\label{HID_V4641Sgr}
\end{figure*}

\end{document}